\documentclass{aa}

\usepackage{graphicx,epsfig}
\usepackage{amssymb,amsmath}
\usepackage{txfonts}
\usepackage{xspace}
\usepackage{natbib}
\usepackage{ulem}
\newcommand{\etal}{{et al.\ }}

\begin{document}

\title{Analysing Globular Cluster observations}
\subtitle{Models and Analysis Tools for Lick/IDS indices}

\author{Thomas Lilly\inst{1} \and Uta Fritze -- v. Alvensleben\inst{2}}

\institute{Institut f\"ur Astrophysik, Friedrich-Hund-Platz 1, 37077 G\"ottingen, Germany\\
           \email{tlilly@astro.physik.uni-goettingen.de}
           \and
           Centre for Astrophysics Research, STRI, University of Hertfordshire,
           College Lane, Hatfield, AL10 9AB, UK\\
           \email{ufritze@star.herts.ac.uk}}

\offprints{T. Lilly}



\date{Received xxx  / Accepted xxx}

\abstract{We have extended our evolutionary synthesis code, {\sc galev}, to include Lick/IDS absorption-line
indices for both simple and composite stellar population models (star clusters and galaxies), using the polynomial
fitting functions of Worthey \etal (1994) and Worthey \& Ottaviani (1997).
We present a mathematically advanced Lick Index Analysis Tool (\emph{LINO}) for the determination of ages and
metallicities of globular clusters (CGs): An extensive grid of {\sc galev} models for the evolution of star
clusters at various metallicities over a Hubble time is compared to observed sets of Lick indices of varying
completeness and precision. A dedicated $\chi^2$ - minimisation procedure selects the best model including $1
\sigma$ uncertainties on age and metallicity. We discuss the age and metallicity sensitivities of individual
indices and show that these sensitivities themselves depend on age and metallicity; thus, we extend Worthey's
(1994) concept of a ``metallicity sensitivity parameter'' for an \emph{old} stellar population at \emph{solar}
metallicity to younger clusters of different metallicities. We find that indices at low metallicity are generally
more age sensitive than at high metallicity.
Our aim is to provide a robust and reliable tool for the interpretation of star cluster spectra becoming available
from 10m class telescopes in a large variety of galaxies -- metal-rich \& metal-poor, starburst, post-burst, and
dynamically young.
We test our analysis tool using observations from various authors for Galactic and M31 GCs, for which reliable age
and metallicity determinations are available in the literature, and discuss in how far the observational
availability of various subsets of Lick indices affects the results. For M31 GCs, we discuss the influence of
non-solar abundance ratios on our results.\\
All models are accessible from our website, \texttt{http://www.astro.physik.uni-goettingen.de/$\tilde{\ }$galev/}.

}

\maketitle
%

\section{Introduction}
\label{Introduction}
In order to understand the formation and evolution of galaxies, one of the essential issues is to reveal their
star formation histories (SFHs). Unfortunately, most galaxies are observable in integrated light only, so that SFH
determinations using the most reliable CMD approach are possible only for a very limited sample of nearby
galaxies.
However, the age and metallicity distributions of star cluster and globular cluster (GC) systems can provide
important clues about the formation and evolutionary history of their parent galaxies. E.g., the violent formation
history of elliptical galaxies as predicted from hierarchical or merger scenarios, is, in fact, more directly
obtained from the age and metallicity distributions of their GC systems than from their integrated spectra that
are always dominated by stars originating from the last major star formation episode. By means of evolutionary
synthesis models, for example, we can show that, using the integrated light of a galaxy's (composite) stellar
content alone, it is impossible to date (and actually to identify) even a very strong starburst if this event took
place more than two or three Gyrs ago (Lilly 2003, Lilly \& Fritze -- von Alvensleben 2005).
Therefore, it is an important first step towards the understanding of the formation and evolution of galaxies to
constrain the age and metallicity distributions of their star cluster systems (for recent reviews see, for example,
Kissler-Patig 2000 and Zepf 1999, 2002), as well as of their stars (see, e.g., Harris \etal 1999, Harris \& Harris
2000, 2002).
Star clusters can be observed one-by-one to fairly high precision in galaxies out to Virgo cluster distances, even
on bright and variable galaxy backgrounds, both in terms of multi-band imaging and in terms of intermediate
resolution spectroscopy.
For young star cluster systems, we have shown that the age and metallicity distributions can be obtained from a
comparision of multi-band imaging data with a grid of model SEDs using the SED Analysis Tool \emph{AnalySED}
(Anders \etal 2004).

Our aim is to extend the analysis of star cluster age and metallicity distributions in terms of parent galaxy
formation histories and scenarios to intermediate age and old star cluster systems.
However, since for all colors the evolution slows down considerably at ages older than about 8 Gyr, even with
several passbands and a long wavelength basis the results are severely uncertain for old GCs; colors -- even when
combining optical and near infrared -- do not allow to completely disentangle the age-metallicity degeneracy (cf.
Anders \etal 2004). Absorption line indices, on the other hand, are a promising tool towards independent and more
precise constraints on ages and metallicities.
Therefore, we present a grid of new evolutionary synthesis models for star clusters including Lick/IDS indices
complementing the broad band colors and spectra of our previous models, and a Lick Index Analysis Tool \emph{LINO}
meant to complement our SED Analysis Tool.
With these two analysis tools, we now possess reasonable procedures for the interpretation of both broad-band color
and spectral index observations.

In an earlier study, we already incorporated a subset of Lick indices into our evolutionary synthesis code {\sc
galev} (Kurth \etal 1999). However, since then the input physics for the code changed considerably; instead of the
older tracks we are now using up-to-date Padova isochrones, which include the thermally pulsing asymptotic giant
branch (TP-AGB) phase of stellar evolution (see Schulz \etal 2002). In this work, we present the integration of the
full set of Lick indices into our code.
Now, our {\sc galev} models consistently describe the time evolution of spectra, broad-band colors, emission lines
and Lick indices for both globular clusters (treated as single-age single-metallicity, i.e. ``simple'' stellar
populations, SSPs) and galaxies (composite stellar populations, CSPs), using the same input physics for all models
(for an exhaustive description of {\sc galev} and its possibilities, as well as for recent extensions of the code
and its input physics, see Schulz \etal 2002, Anders \& Fritze -- v. Alvensleben 2003, and Bicker \etal 2004).

A recent publication (Proctor \etal 2004) also presented an analysis tool for Lick indices using a
$\chi^2$-approach. However, they do not provide any confidence intervals for their best fitting models. In this
respect, our new tool extends their approach. A drawback of our models is that, at the present stage, they do not
account for variations in $\alpha$-enhancement, as Proctor \etal (2004) do. However, our analysis tool \emph{LINO}
is easily applicably to any available set of absortion line indices.\\

In Section \ref{Galev}, we recall the basic definitions of Lick indices, and describe how we synthesize them in
our models; we also address non-solar abundance ratios. Some examples for SSP model indices are presented and
briefly confronted with observations.
In Section \ref{Sensitivity}, Worthey's (1994) ``metallicity sensitivity parameter'' is discussed and extended
from old stellar populations to stellar populations of all ages.
Section \ref{Analysis} describes and tests our new Lick Index Analysis Tool; Galactic and M31 globular cluster
observations are analysed and compared with results (taken from literature) from reliable CMD analysis, and from
index analyses using models with varying $\alpha$-enhancements, respectively.
Section \ref{Conclusions} summarizes the results and provides an outlook.\\

\section{Models and input physics}
\label{Galev}

\subsection{Evolutionary synthesis of Lick indices}
\label{synthesis}

Lick indices are relatively broad spectral features, and robust to measure.
They are named after the most prominent absorption line in the
respective index's passband. However, this does not necessarily mean
that a certain index's strength is exclusively or even dominantly due to line(s) of this element.
(see, e.g., Tripicco \& Bell 1995). Beyond the fact that more than one
line can be present in the index's passband, strong lines in the
pseudo-continua can also affect the index strength.
Most indices are given in units of their equivalent width (EW)
measured in \AA:
\begin{equation}
   \label{Lick_EW}
   \text{EW[\AA]} = \int_{\lambda_1}^{\lambda_2} \frac{F_C(\lambda) - F_I(\lambda)}{F_C(\lambda)} \ {\rm d}\lambda \ ,
\end{equation}
whereas index strengths of broad molecular lines are given in
magnitudes:
\begin{equation}
   \label{Lick_mag}
   \text{I[mag]} = -2.5 \ \log \left[ \left(\frac{1}{\lambda_1-\lambda_2}\right) 
           \int_{\lambda_1}^{\lambda_2} \frac{F_I(\lambda)}{F_C(\lambda)} \ {\rm d}\lambda \right] .
\end{equation}

$F_I(\lambda)$ is the flux in the index covering the wavelength range
between $\lambda_1$ and $\lambda_2$; $F_C(\lambda)$ is the continuum
flux defined by two ``pseudo-continua'' flanking the central index
passband.

There are currently 25 Lick indices, all within the optical wavelength
range:
H$\delta_A$, H$\gamma_A$, H$\delta_F$, H$\gamma_F$, 
CN$_1$, CN$_2$, Ca4227, G4300, Fe4383, Ca4455, Fe4531, Fe4668,
H$\beta$, Fe5015, Mg$_1$, Mg$_2$, Mg\textsl{b}, Fe5270, Fe5335,
Fe4506, Fe5709, Fe5782, Na D, TiO$_1$, and TiO$_2$. For a full
description and all index definitions, see Trager \etal (1998) and
references therein.\\

As the basis for our evolutionary synthesis models we employ the polynomial fitting functions of Worthey
\etal (1994) and Worthey \& Ottaviani (1997), which give Lick index strenghts of individual stars as a function of their
effective temperature $T_{\rm eff}$, surface gravity $g$, and metallicity [Fe/H]. Worthey \etal have calibrated
their fitting functions empirically using solar-neighbourhood stars.

Model \emph{uncertainties} are calculated as follows (Worthey 2004):
\begin{equation}
   \label{err}
   \sigma_{\rm model} = \frac{\sigma_{\rm star} \times {\rm RMS_{fit}}}{\sqrt{N}}
\end{equation}
with $\sigma_{\rm star}$ being the typical rms error per observation for the calibration stars and ${\rm
RMS_{fit}}$ being the residual rms of the fitting functions (both values are given in Worthey \etal (1994) and
Worthey \& Ottaviani (1997), respectively). N is the number of stars in the ``neighbourhood'' of the fitting
functions in the $T_{\rm eff}, g$, [Fe/H] space, which is typically of the order of 25.
Note that this approach is only an approximation; the real model error is most likely a strong function of $T_{\rm
eff}, g$, and [Fe/H].

Other input physics of our models include the theoretical spectral library from Lejeune \etal (1997, 1998) as well
as theoretical isochrones from the Padova group for $Z$=0.0004, 0.004, 0.008, 0.02 and 0.05 (cf. Bertelli \etal
1994), and for $Z$=0.0001 (cf. Girardi \etal 1996); recent versions of these isochrones include the TP-AGB phase
of stellar evolution (not presented in the referenced papers) which is important for intermediate age stellar
populations (cf. Schulz \etal 2002). We assume a standard Salpeter (1955) initial mass function (IMF) from 0.15 to
about 70 M$_\odot$; lowest mass stars (M$_\odot < 0.6$) are taken from Chabrier and Baraffe (1997) (cf. Schulz
\etal 2002 for details). Throughout this paper, we identify the metallicity $Z$ with [Fe/H] and define [Fe/H] $:=
\log(Z/Z_{\odot})$.\\

To calculate the time evolution of Lick indices for SSP or galaxy
models, we follow four steps:
\begin{enumerate}
  \item We use the values for $T_{\rm eff},\ g,$ and [Fe/H] given
  (directly or indirectly) by the isochrones to compute the index
  strength EW$_{\rm star}$ or I$_{\rm star}$ for each star along the isochrones.
  \item A spectrum is assigned to each star on a given isochrone and
  used to compute its continuum flux $F_C$\footnote{In view of the resolution
  of our spectral library, these values are not very accurate;
  however, since $F_C$ is merely an additional weighting
  factor for the integration routine, this does not affect the final
  results.}.
  \item For each isochrone, the index strengths are integrated over
  all stellar masses $m$ (after transformation of the index strengths into
  fluxes), weighted by the IMF (using a weighting factor $w$):
  \begin{equation}
     \label{Lick_integration}
     \text{EW$_{\rm SSP}$} = (\lambda_2-\lambda_1) \cdot \left( 1 - \frac{\sum_{m} (F_I \cdot
     w)}{\sum_{m} (F_C \cdot w)} \right) \ ,
  \end{equation}
   with $F_I$ being a function of EW$_{\rm star}$ and $F_C$:
  \begin{equation}
     \label{Lick_FluxI}
     F_I = F_C \cdot \left( 1 - \frac{\text{EW}_{\rm star}}{\lambda_2-\lambda_1} \right).
  \end{equation}
\end{enumerate}
The result is a grid of SSP models for all available isochrones, i.e.,
for 50 ages between 4 Myr and 20 Gyr, and the 6 metallicities given
above.
\begin{itemize}
  \item[4.] For each time step in the computation of a stellar
  population model, our evolutionary synthesis code {\sc galev} gives
  the contribution of each isochrone to the total population.\\
  To obtain galaxy model indices (or a better age resolution for SSP
  models), we integrate our grid of SSP models using equations
  (\ref{Lick_integration}) and (\ref{Lick_FluxI}) again, but with $w$
  being the isochrone contribution as a new weighting factor (now doing
  the summation over all isochrones instead of all masses), $F_C$
  being the integrated continuum flux level for each isochrone, and
  using EW$_{\rm SSP}$ instead of EW$_{\rm star}$.
\end{itemize}

Following this way, we have computed a large grid of SSP models, consisting of 6 metallicities and 4000 ages from 4
Myr to 16 Gyr in steps of 4 Myr; each point of the model grid consists of all 25 Lick indices currently available.

\subsection{Non-solar abundance ratios}
\label{alphaenhancement}
Abundance ratios reflect the relation between the characteristic
time scale of star formation and the time scales for the release of,
e.g., SNe II products (Mg and other $\alpha$-elements), SNe Ia products
(Fe), or nucleosynthetic products from intermediate-mass stars (N).
Galaxies with different SFHs 
will hence be characterised by different distributions of stellar
abundances ratios.
This means, the Galactic relation between abundance ratios and
metallicity (Edvardsson \etal 1993, Pagel \& Tautvai\v{s}ien\.e 1995)
is not necessarily valid for galaxies of different types and formation
histories.
Empirical index calibrations based on Galactic stars,
like the fitting functions from Worthey \etal (1994) and Worthey \&
Ottaviani (1997) applied in this work, are based on the implicit
inclusion of the Galactic relation between abundance ratios and
metallicity.

A lot of work has been done in the last years to study the impact of
$\alpha$-enhancement on stellar population models and their applications:
E.g., based mainly on the work of Tripicco \& Bell (1995) and Trager \etal
(2000a), Thomas \etal (2003, 2004) presented SSP models of Lick indices
with variable abundance ratios that are corrected for the bias mentioned
above, providing for the first time well-defined [$\alpha$/Fe] ratios at
all metallicities. 
The impact of these new models on age and metallicity estimates of early
type galaxies is investigated in detail by Maraston \etal (2003),
Thomas \& Maraston (2003), Thomas \etal (2004), as well as by Trager
\etal (2000a,b), among others.\\

However, since our purpose is to present consistently computed models
for spectra, colors, emission lines \emph{and} Lick indices for both SSPs and
CSPs, a consistent attempt to allow our evolutionary synthesis code {\sc
galev} to account for arbitrary abundance ratios would have to be based
on stellar evolutionary tracks or isochrones, detailed nucleosynthetic
stellar yields, and model atmospheres for various abundance ratios.
Since both consistent and complete datasets of this kind are not yet
available (though first sets of evolutionary tracks for stars with
enhanced [$\alpha$/Fe] ratios were presented by Salasnich \etal 2000 and
Kim \etal 2002), at the present stage our models do \emph{not} explicitly allow for
variations in $\alpha$-enhancement.
This is an important caveat to be kept in mind for the interpretation of
extragalactic GC populations. We think that the extensive
studies on non-solar abundance ratios cited above will allow to estimate
the impact of this caveat on our results.

However, in Section \ref{alphatest} we show that our method is robust enough to give very good age and metallicity
determinations for GCs even without using $\alpha$-enhanced models.

\subsection{SSP model indices: Some examples}
\label{models}
\begin{figure*}
   \begin{center}
   \includegraphics[height=0.49\linewidth,angle=270]{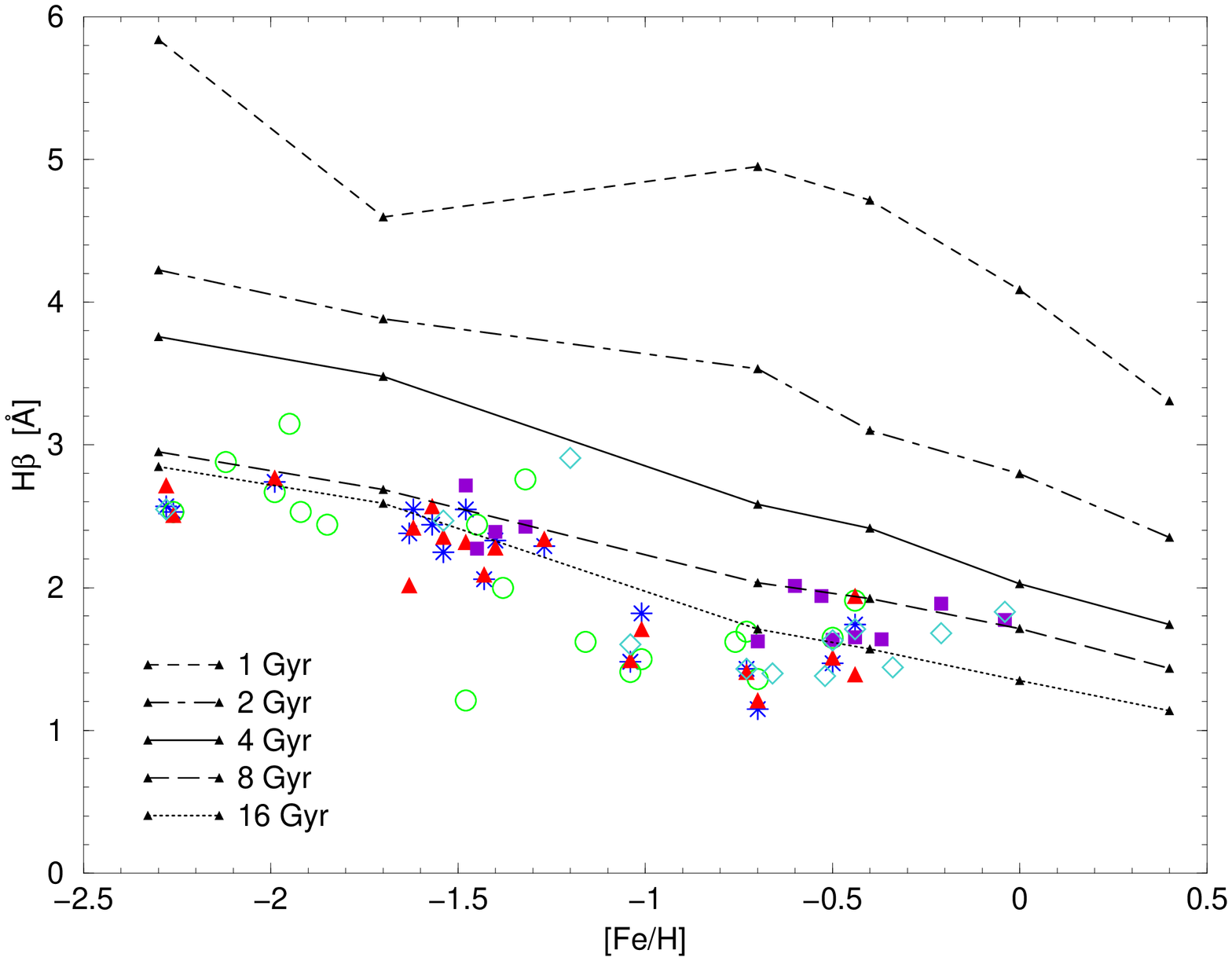}
   \includegraphics[height=0.49\linewidth,angle=270]{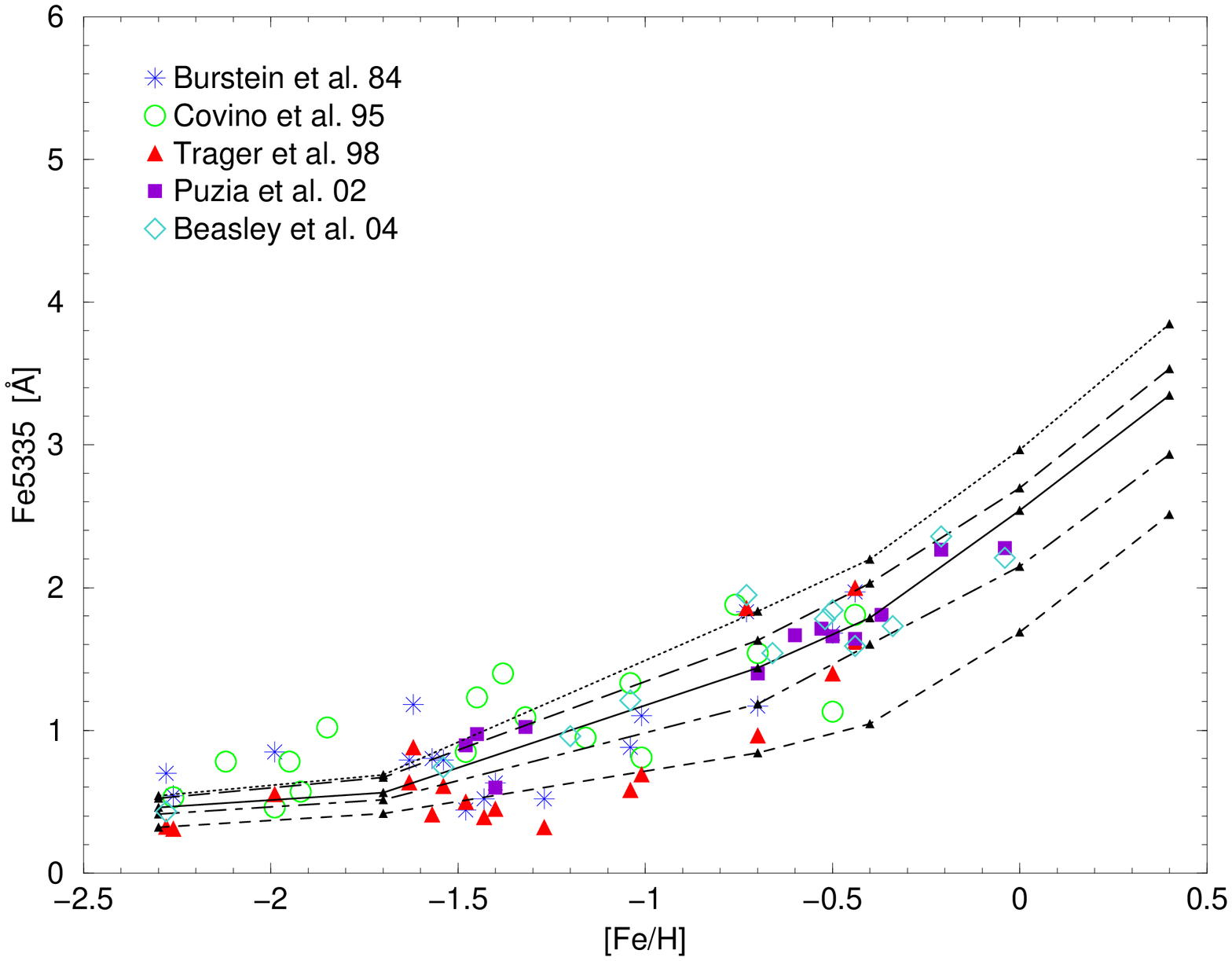}
   \caption{Indices H$\beta$ \emph{(left)} and Fe5335 \emph{(right)}
   versus metallicity for 5 different ages. Also shown are Galactic GC
   observations from various authors as indicated in the right-hand
   panel; GC metallicities are taken from Harris (1996, revision
   Feb. 2003). A typical measurement error is about 0.2 \AA.}
   \label{abb.lickdata}
   \end{center}
\end{figure*}
\begin{figure*}
   \begin{center}
   \includegraphics[height=0.49\linewidth,angle=270]{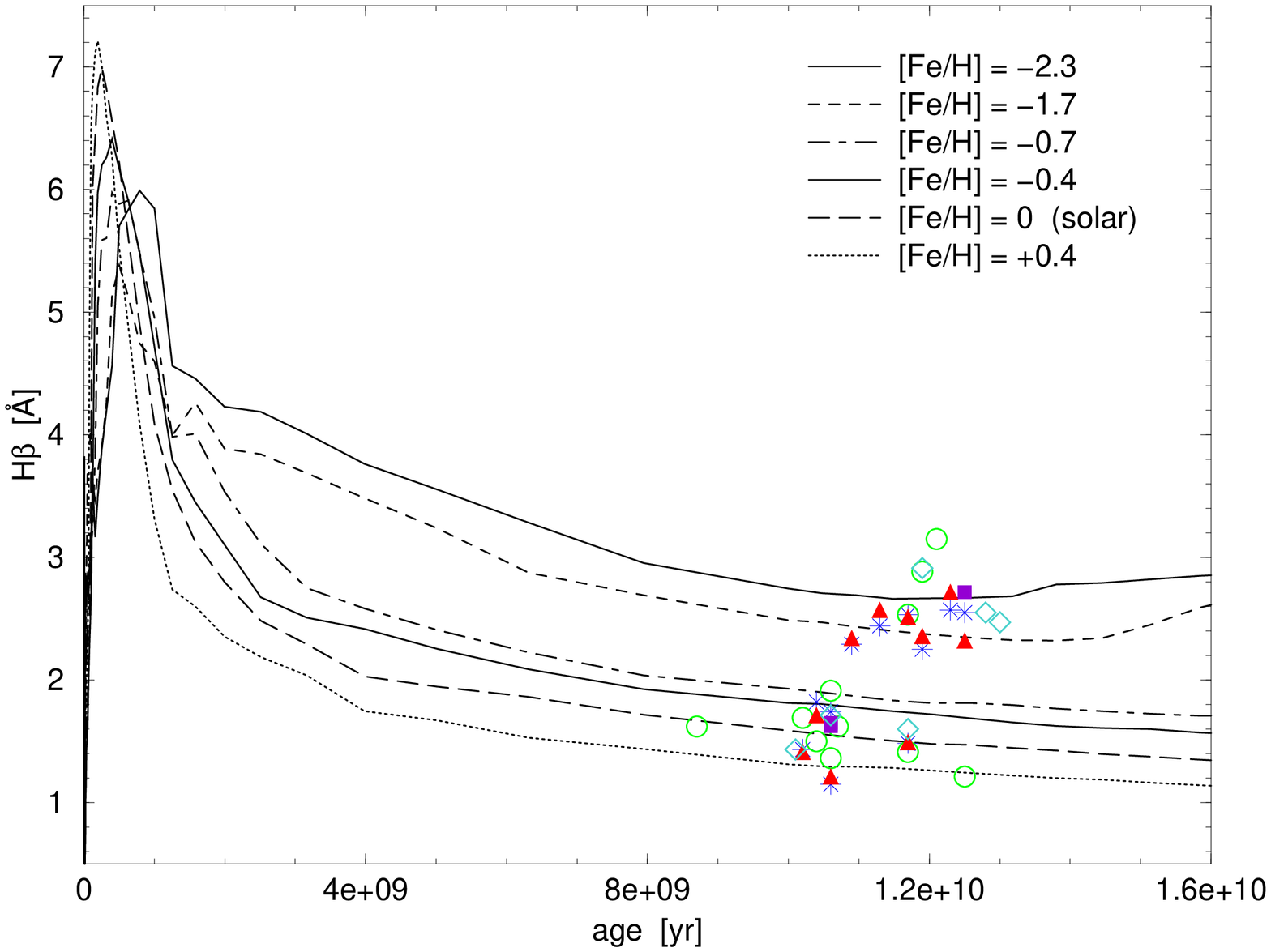}
   \includegraphics[height=0.49\linewidth,angle=270]{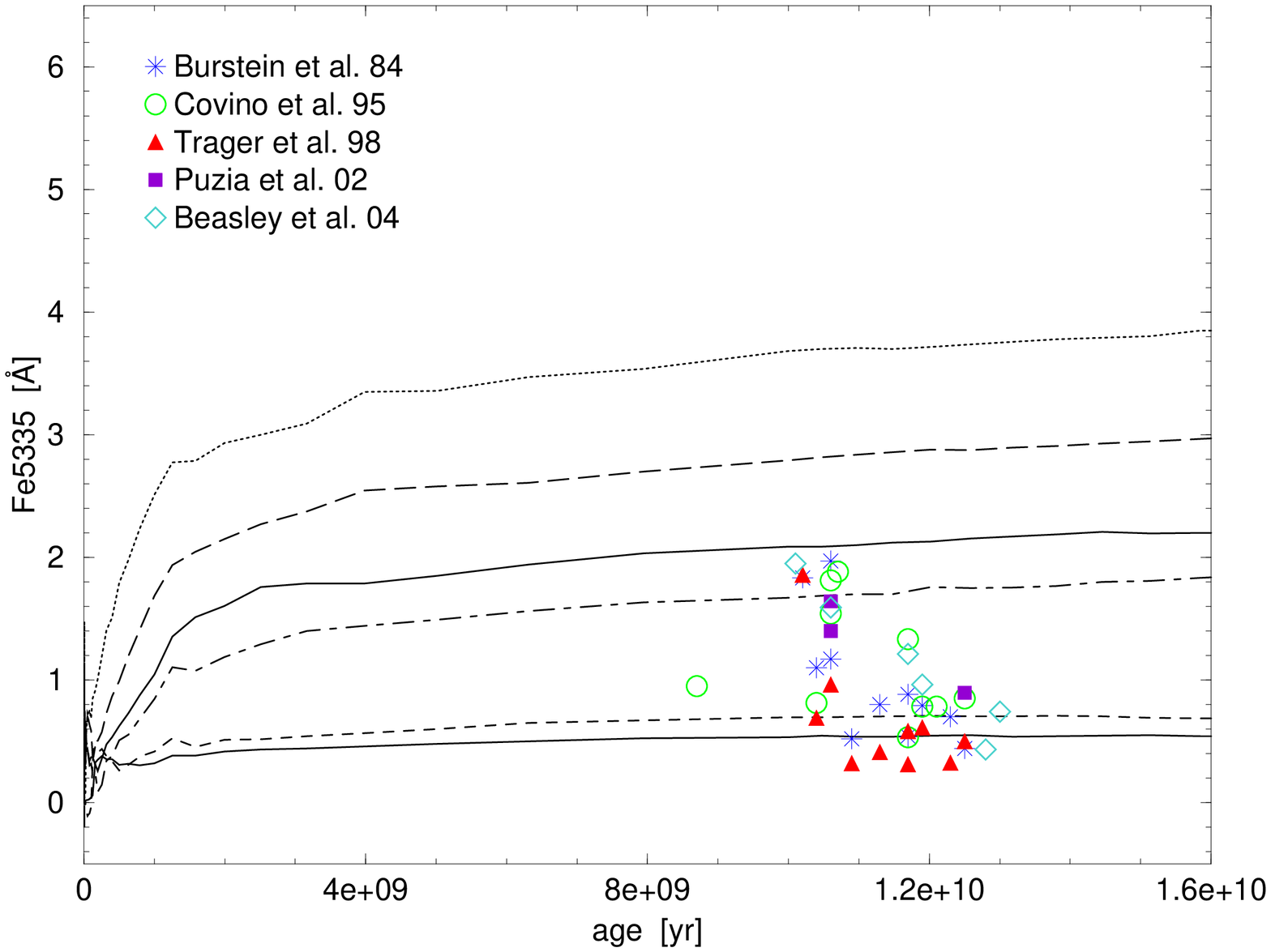}
   \caption{Indices H$\beta$ \emph{(left)} and Fe5335 \emph{(right)}
   versus age for 6 different metallicities. Also shown are Galactic
   GC observations from various authors as indicated in the right-hand
   panel; GC age determinations are taken from Salaris \& Weiss
   (2002).}
   \label{abb.sspevolution}
   \end{center}
\end{figure*}
In Figs. \ref{abb.lickdata} and \ref{abb.sspevolution}, we show the time evolution and metallicity dependence of
the indices H$\beta$ and Fe5335 in our new SSP models, and compare them with index measurements of Galactic GCs
that are plotted against reliable age and metallicity estimates, respectively.

In particular, in Fig. \ref{abb.lickdata} we confront SSP models for five
ages between 1 and 16 Gyr with Galactic GC observations by Burstein
\etal (1984; 17 clusters), Covino \etal (1995; 17 clusters), Trager
\etal (1998; 18 clusters), Puzia \etal (2002;
12 clusters) and Beasley \etal (2004; 12 clusters). Note that
some clusters were repeatedly observed, so more than one data point in
the figure can refer to the same cluster. The metallicities
are taken from Harris (1996, revision Feb. 2003).
In Fig. \ref{abb.sspevolution} we show the time evolution of the model indices for all six metallicities, and
confront them with Galactic GC observations (taken from the same references as in Fig. \ref{abb.lickdata}). The GC
age determinations are based on CMD fits and taken from Salaris \& Weiss (2002)\footnote{Note that they only cover
a subsample of the observations shown in Fig.\ref{abb.lickdata}: 11 clusters of the Burstein \etal (1984) sample,
10 of the Covino \etal (1995) sample,  10 of the Trager \etal (1998) sample, only 3 of the Puzia \etal (2002)
sample, and 6 clusters of the Beasley \etal (2004) sample.}.
Over the range of Galactic GC ages and metallicities (i.e., ages older than about 8 Gyr and metallicities lower
than solar in most cases), a sufficient agreement is observed between models and data in the sense that the data
lie within the range of the model grid; we also checked this for other indices (not plotted).

However, the plots also demonstrate how difficult it would be to interpret the indices in terms of classical
index-index plots. Actually, Fig. \ref{abb.sspevolution} seems to show apparent inconsistencies. So, some clusters
in Fig. \ref{abb.sspevolution} have metallicities up to [Fe/H]=+0.4 when compared with models for the age
sensitive index H$\beta$, whereas, when compared with models for the metallicity sensitive Fe5335 index, all
clusters have metallicities lower than [Fe/H]=-0.4.
We cannot decide at this point to what degree these inconsistencies are due to problems in the \emph{models} or
the calibrations they are based on or due to badly calibrated \emph{observations}; however, our new Lick Index
Analysis Tool nevertheless gives surprisingly robust age and, particularly, metallicity determinations for the
same set of cluster observations (cf. Sect. \ref{tests}).

\section{Index sensitivities}
\label{Sensitivity}
It is well known that different indices have varying sensitivities to age and/or metallicity. To quantify this,
Worthey (1994) introduced a ``metallicity sensitivity parameter'' that gives a hint on how sensitive a given index
is with respect to changes in age and metallicity.
This parameter is defined as the ratio of the percentage change in Z to the percentage change in age (so influences
of possible age-metallicity degeneracies are implicitely included), with large numbers indicating greater
metallicity sensitivity:
\begin{equation}
   \label{eq.s_parameter}
   S = \left( \frac{\Delta I_Z}{\Delta Z/Z} \right) \left \slash \left( \frac{\Delta I_{age}}{\Delta age/age} \right) \right .
\end{equation}
Using his SSP models, Worthey (1994) chose a 12 Gyr solar metallicity (Z=0.017) model as the zero point for the
sensitivity parameters, the $\Delta$'s referring to ``neighbouring'' models, in this case models with age = 8/17
Gyr (i.e, $\Delta$age = 4/5 Gyr) and Z = 0.01/0.03 (i.e., $\Delta$Z = 0.007/0.013)\footnote{Ideally, S should be
relatively independent of the exact values of the $\Delta$Z and $\Delta$age chosen, as long as they are not too
large.}; the main numerator/denominator in Eq. \ref{eq.s_parameter} is averaged using both $\Delta$'s before
computing the fraction.\\

\begin{table}
   \caption{Metallicity sensitivity parameters for different zero points. Low numbers indicate high
   age-sensitivity. Values given in brackets are not reliable (see text).}
   \label{table.sensitivities}
\centering
\begin{tabular}{c c c c c c}
        \hline\hline
          & \textsc{Worthey} & \multicolumn{2}{c}{GALEV 12 Gyr} & \multicolumn{2}{c}{GALEV 4 Gyr}\\
          &  & Z = 0.02 & 0.0004 & 0.02 & 0.0004\\
        \hline
	CN$_1$		& 1.9	&  1.5   &  1.1   &  1.1   &  0.1\\
	CN$_2$		& 2.1	&  1.5   &  0.2   &  1.2   &  0.1\\
	Ca4227		& 1.5	&  1.1   & (0.4)  &  1.0   &  0.1\\
	G4300		& 1.0	&  1.0   &  0.2   &  0.8   &  0.1\\
	Fe4383		& 1.9	&  1.9   &  0.3   &  1.3   &  0.2\\
	Ca4455		& 2.0	&  1.7   &  0.7   &  1.5   &  0.3\\
	Fe4531		& 1.9	&  1.7   &  0.6   &  1.4   &  0.2\\
	Fe4668		& 4.9	& (3.5)  & (0.9)  &  2.4   &  0.9\\
	H$\beta$	& 0.6	&  0.6   &  0.1   &  0.5   &  0.1\\
	Fe5015		& 4.0	& (2.3)  & (1.3)  &  2.1   &  0.4\\
	Mg$_1$		& 1.8	&  1.7   & (2.2)  &  1.4   &  2.0\\
	Mg$_2$		& 1.8	&  1.5   & (1.8)  &  1.2   &  0.5\\
	Mg\textsl{b}	& 1.7	&  1.4   & (0.7)  &  1.0   &  0.3\\
	Fe5270		& 2.3	&  2.0   & (0.7)  &  1.6   &  0.3\\
	Fe5335		& 2.8	&  2.7   & (1.3)  &  2.0   &  0.4\\
	Fe5406		& 2.5	& (2.6)  & (2.3)  &  1.8   &  0.6\\
	Fe5709		& 6.5	& (8.5)  & (1.7)  &  2.6   & (1.2)\\
	Fe5782		& 5.1	& (5.9)  & (1.4)  &  2.5   & (1.0)\\
	Na D		& 2.1	&  1.9   & (1.2)  &  1.9   &  0.6\\
	TiO$_1$		& 1.5	&  0.9   &  0.7   & (1.4)  & (5.5)\\
	TiO$_2$		& 2.5	&  1.3   &  0.9   & (1.6)  & (8.6)\\
	H$\delta_A$	& 1.1	&  1.0   &  0.3   &  0.8   &  0.1\\
	H$\gamma_A$	& 1.0	&  1.0   &  0.2   &  0.8   &  0.1\\
	H$\delta_F$	& 0.9	&  0.9   &  0.1   &  0.7   &  0.1\\
	H$\gamma_F$	& 0.8	&  0.8   &  0.2   &  0.7   &  0.1\\
        \hline
\end{tabular}
\end{table}

In Table \ref{table.sensitivities}, we reprint the metallicity sensitivity parameters given by Worthey (1994) and
Worthey \& Ottaviani (1997), and confront them with parameters computed using our own models. Extending Worthey's
approach, we compute parameters for four different combinations of zero points, using high (Z = 0.02) and low
(Z = 0.0004) metallicities, and high (12 Gyr) and intermediate (4 Gyr) ages.

\begin{table}
   \caption{``Neighbouring models'' in terms of metallicity for Z=0.02 and Z=0.0004 (\emph{top}) and in terms of
   age for the zero points 12 Gyr and 4 Gyr (\emph{bottom}) used to compute the metallicity sensitivities given in
   Table \ref{table.sensitivities}. Brackets give the corresponding $\Delta$Z and $\Delta$age, respectively.}
   \label{table.deltas}
\centering
\begin{tabular}{c c c c}
        \hline\hline
          \multicolumn{2}{c}{Z=0.02} & \multicolumn{2}{c}{Z=0.0004}\\
        \hline
          0.008(0.012) & 0.05(0.03) & 0.0001(0.0003) & 0.004(0.0036)\\
        \hline
\end{tabular}
\vspace{0.5cm}

\begin{tabular}{c c c c}
        \hline\hline
          \multicolumn{2}{c}{12 Gyr} & \multicolumn{2}{c}{4 Gyr}\\
        \hline
          11.0(1.0) & 13.2(1.2) & 3.2(0.8) & 5.0(1.0)\\
          10.5(1.5) & 13.8(1.8) & 2.5(1.5) & 6.3(2.3)\\
        \hline
\end{tabular}
\end{table}

Worthey's parameters are relatively well reproduced by models with a similar combination of zero points, i.e. for
old (12 Gyr) and solar metallicity SSPs; however, S is not totally independent of the $\Delta$Z and $\Delta$age
chosen; it can be very sensitive to the exact evolution of the model index, mainly in age-metallicity space
regions where the slope of the index does not evolve very smoothly (for example, a very high value of the
parameter can also mean that, due to a small ``bump'' in the time evolution of the index, $\Delta I_{age}$ is near
zero; in this case, S is worthless).
The two zero points for both age and metallicity and their ``neighbouring models'' we use for the computations
are given in Table \ref{table.deltas}; for both age zero points, we chose two sets of neighbouring models
and averaged the final parameters.
In order to check the reliability of our results, we also computed parameters for values of age and $\Delta$age not
given in the table. If the results for different $\Delta$age's (or slightly different zero points) differ strongly,
we classify the parameter as uncertain (indicated by brackets in Table \ref{table.sensitivities}).

The ``ranking'' of indices in terms of sensitivity is, with some exceptions, unaffected by changes in the zero
points.

However, contrary to our expectations, at solar metallicity the age-sensitivity of Lick indices is only
slightly higher for an intermediate age model compared to the 12 Gyr model; for low-metallicity SSPs, the effect
is more pronounced. Most important, we find the result that for models at low metallicity, indices are generally
much more age sensitive than for models at high metallicity, especially for age sensitive indices like G4300 or
Balmer line indices. This means, indices of old, low metallicity GCs can be more sensitive to age than indices
of GCs with high metallicity and intermediate age.
This is of special interest for any analysis of GC systems involving intermediate age GCs (e.g., in merger
remnants), since secondary GC populations with intermediate ages are generally expected to have higher
metallicities than ``normal'' old and metal-poor populations.\\

\begin{figure*}
   \begin{center}

   \includegraphics[width=0.29\linewidth,angle=270]{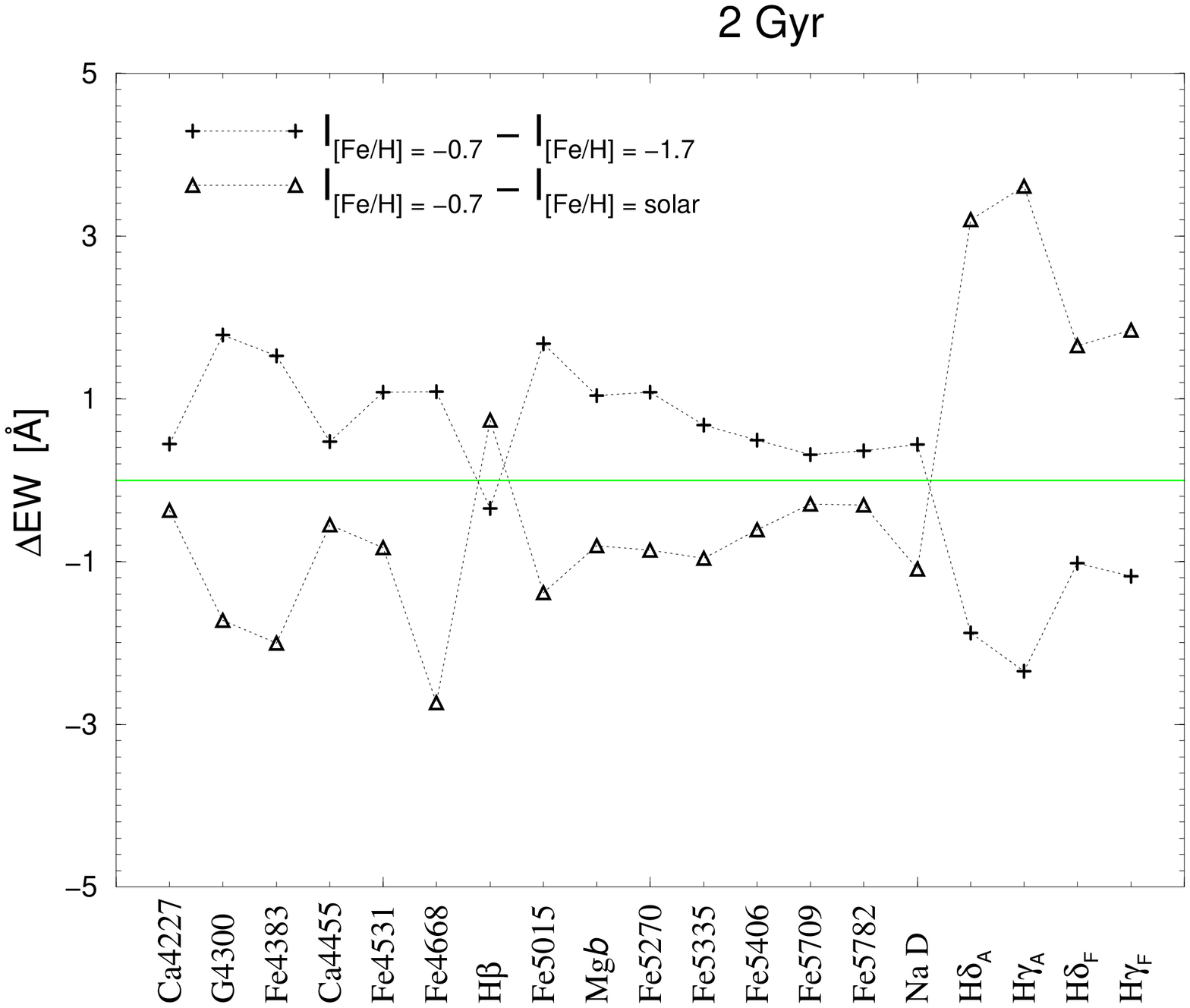}
   \includegraphics[width=0.29\linewidth,angle=270]{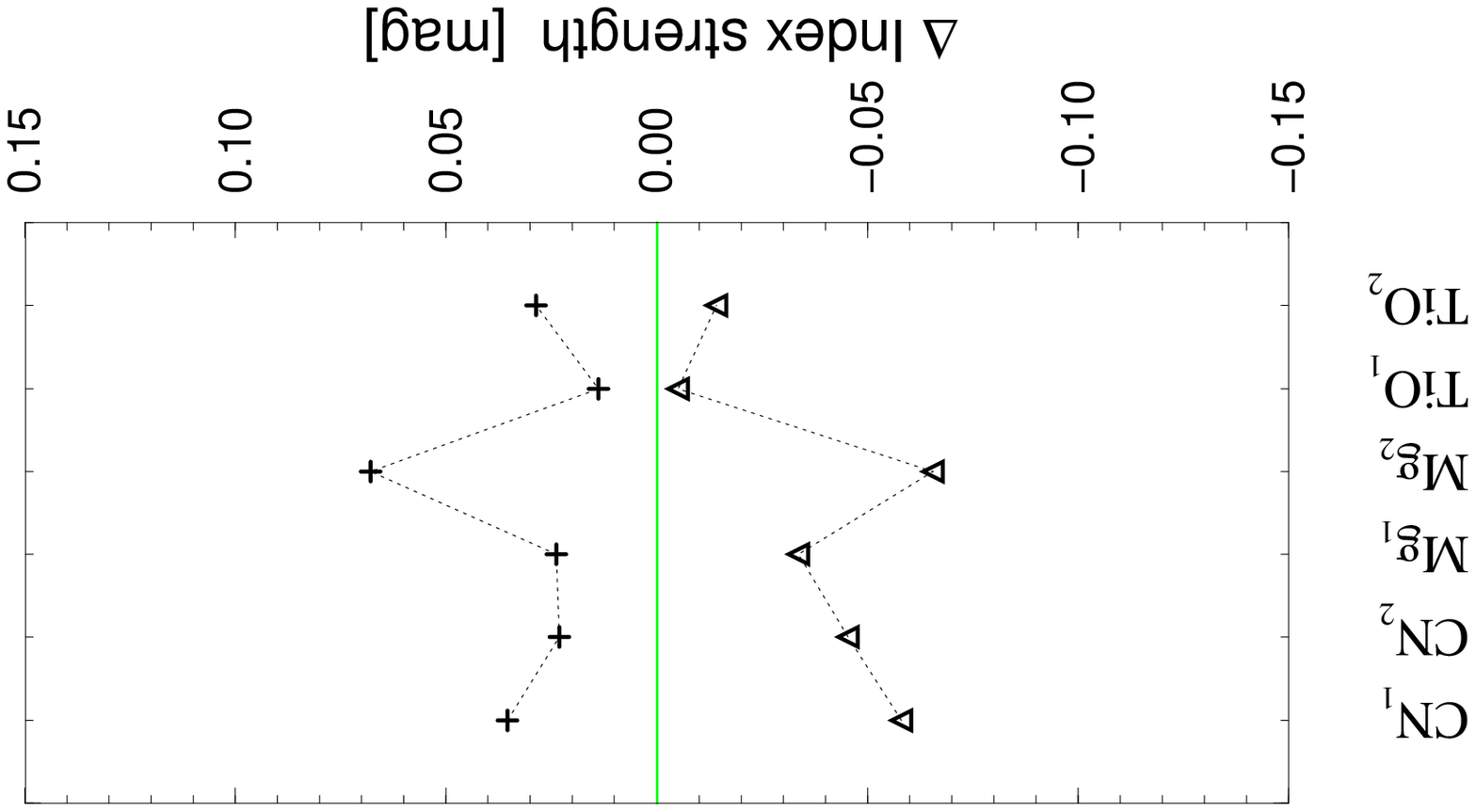}
   \includegraphics[width=0.29\linewidth,angle=270]{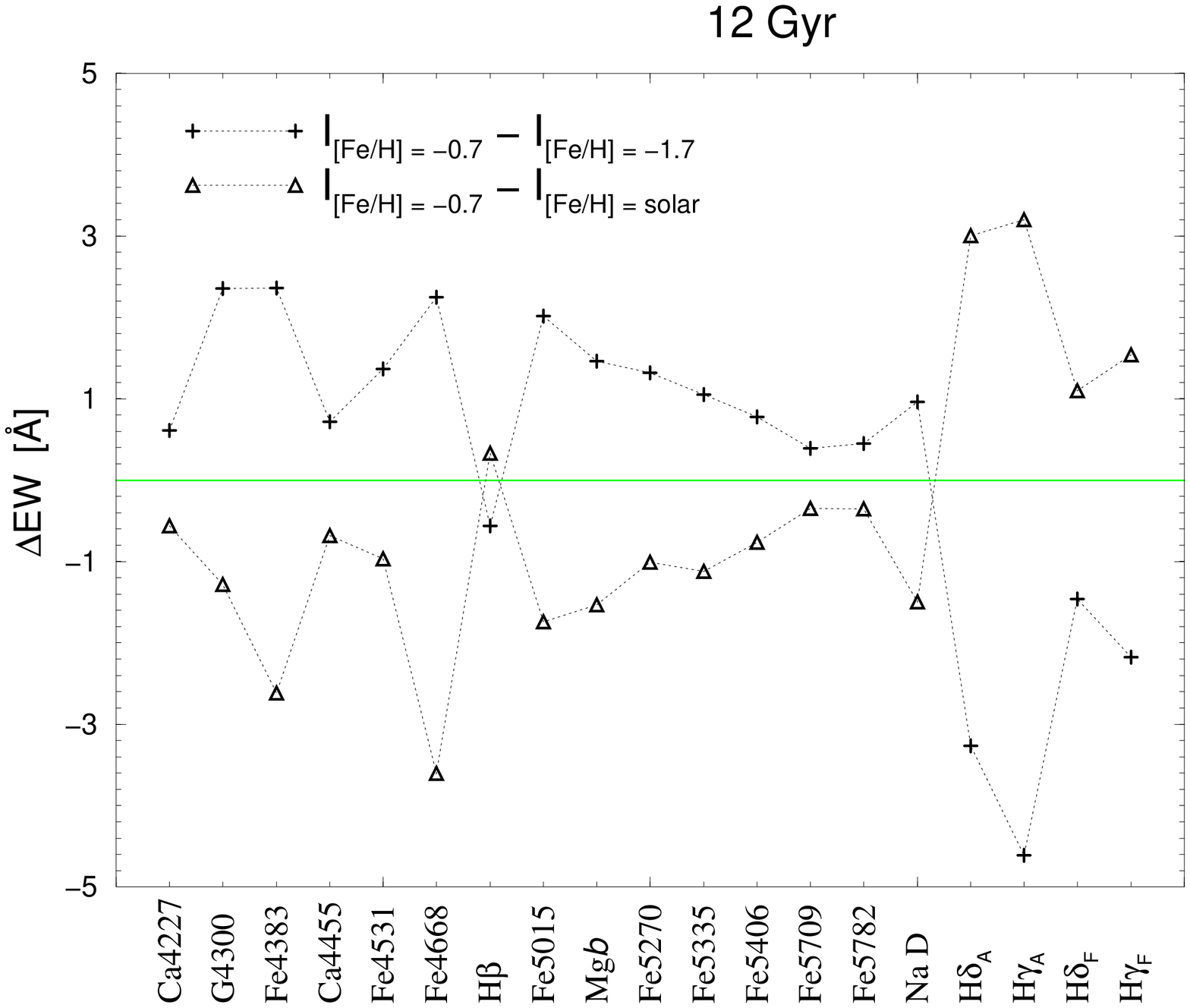}
   \includegraphics[width=0.29\linewidth,angle=270]{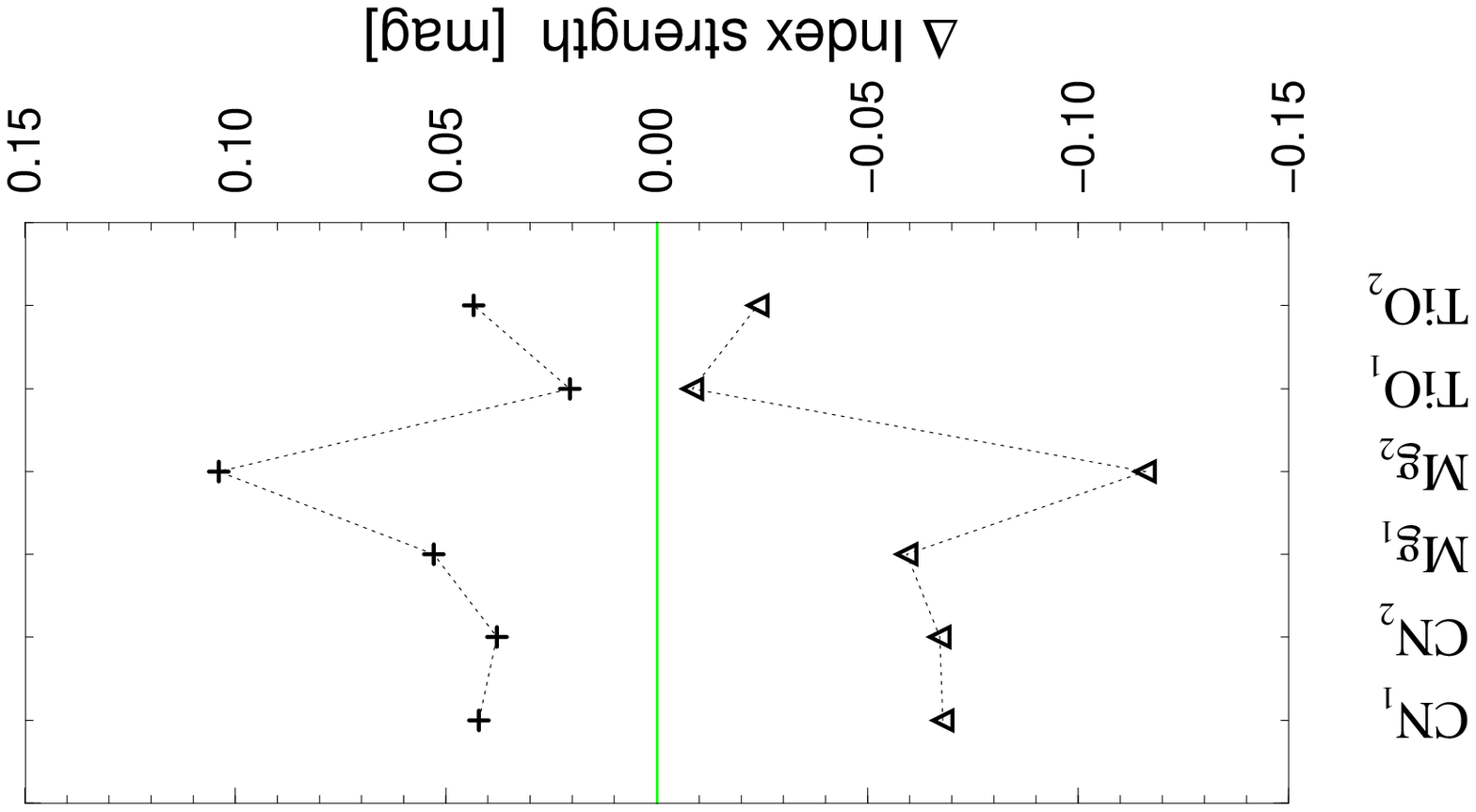}

   \includegraphics[width=0.29\linewidth,angle=270]{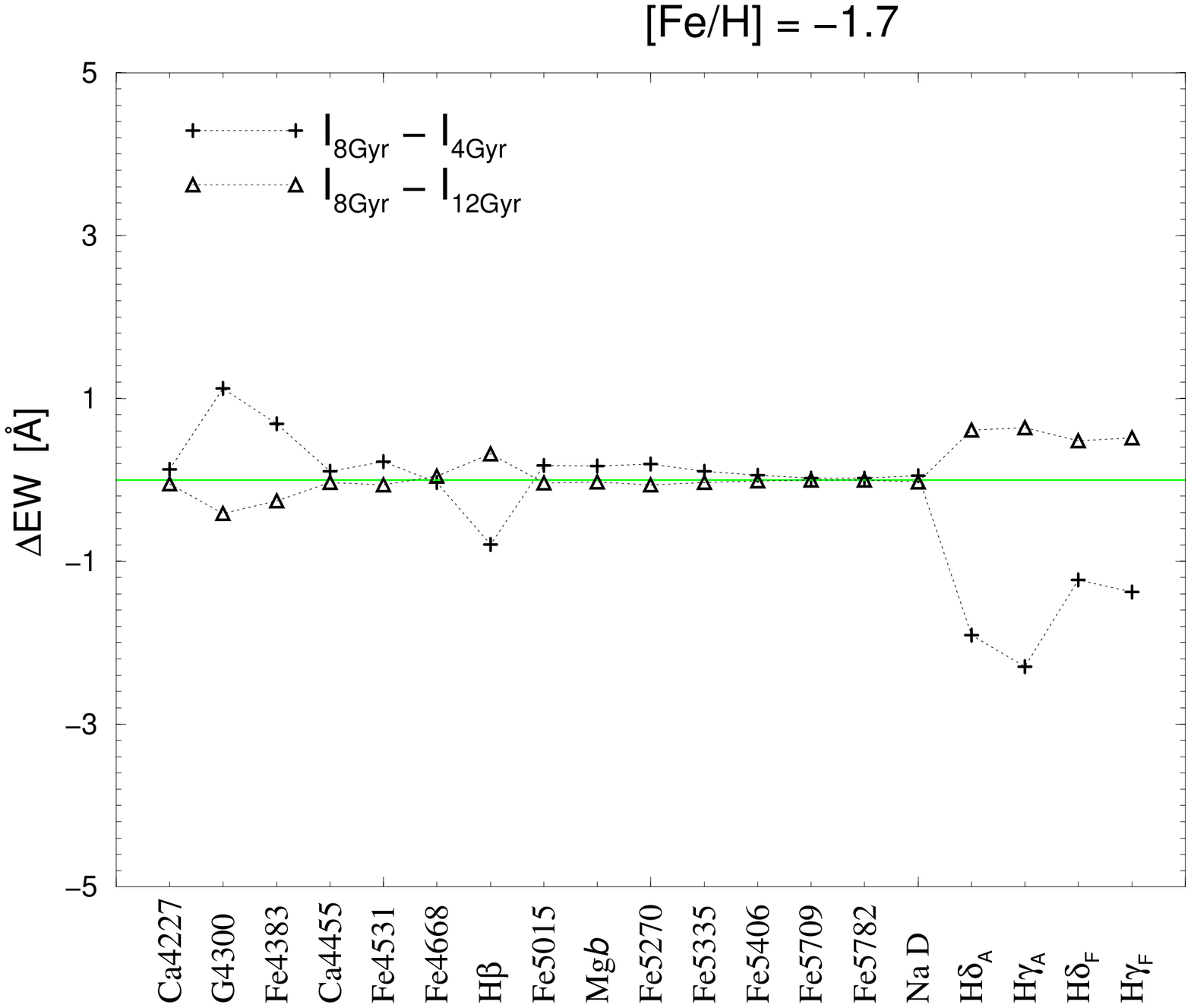}
   \includegraphics[width=0.29\linewidth,angle=270]{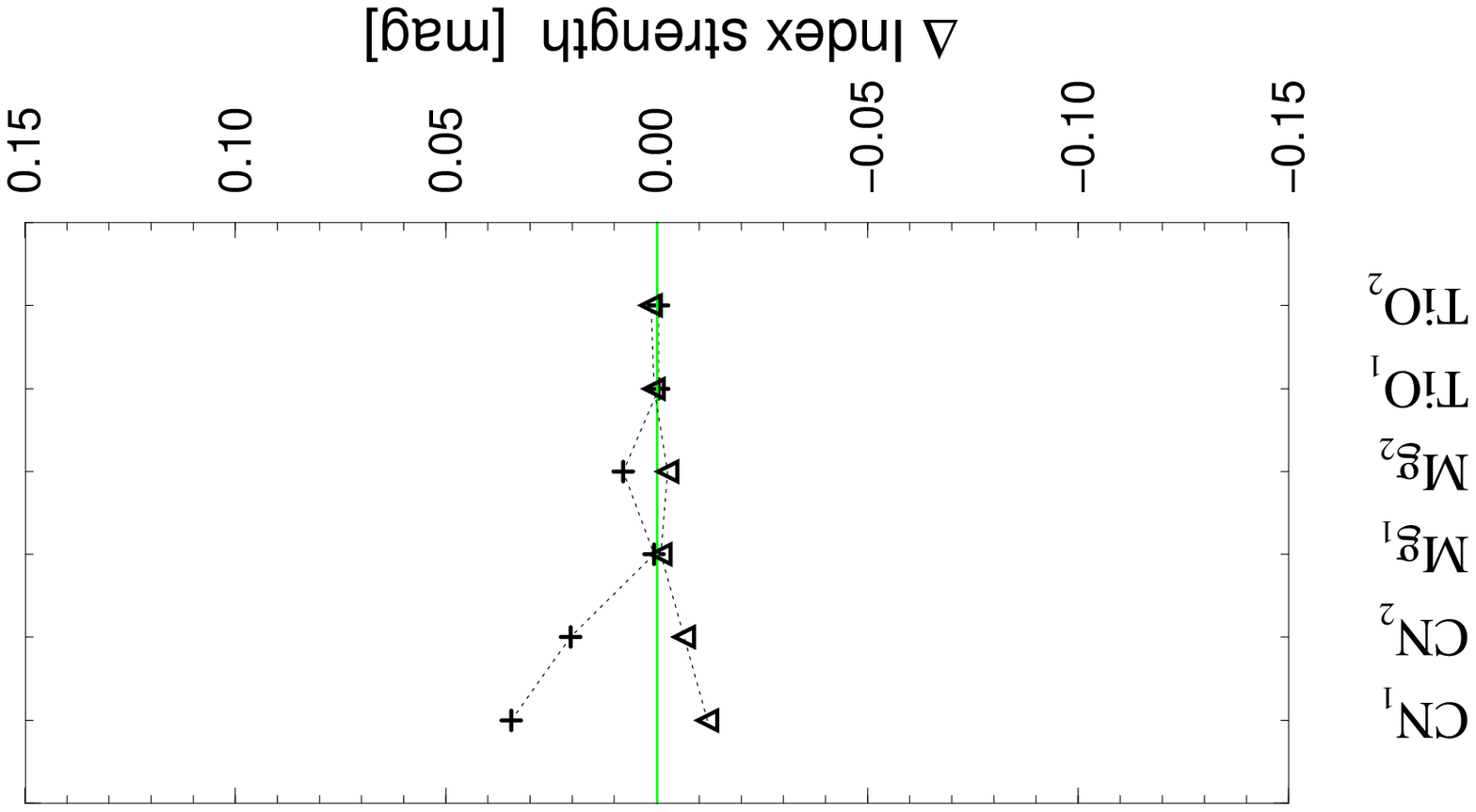}
   \includegraphics[width=0.29\linewidth,angle=270]{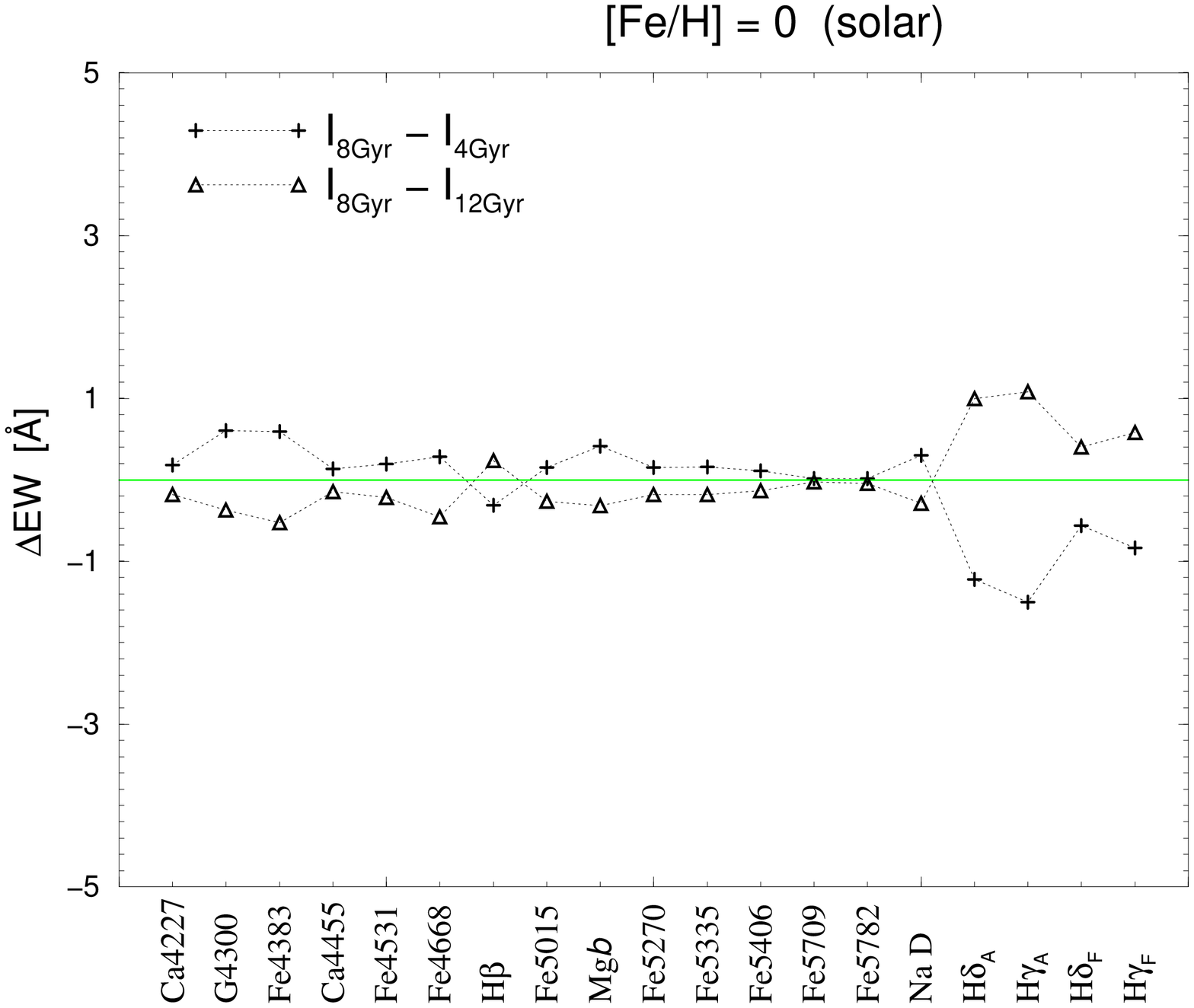}
   \includegraphics[width=0.29\linewidth,angle=270]{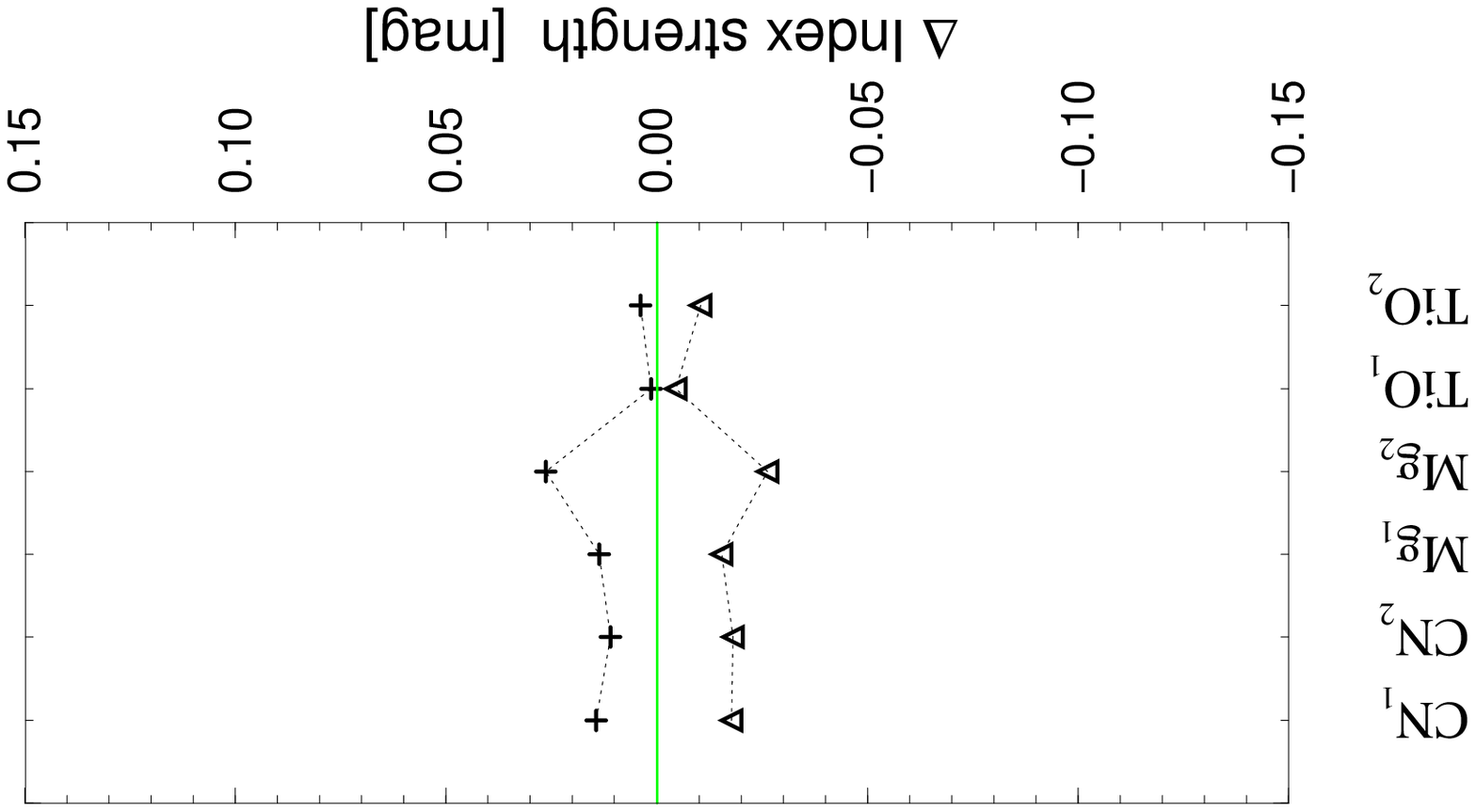}

   \caption{Absolute differences of index strengths for old and young SSP models for changing metallicity
   (\emph{top}), and for metall-rich and metall-poor SSP models for changing age (\emph{bottom}). The dotted lines
   are just for presentation.}
   \label{abb.sensitivities}

   \end{center}
\end{figure*}
Given the limited accuracy of any index measurement, in practice the usefulness of an index to determine age or
metallicity does not only depend on the \emph{relative} change in index strength for changing Z or age as it is
given by S but also on the \emph{absolute} change in index strength.

Therefore, in Figure \ref{abb.sensitivities} we show the absolute differences of index strengths for old (12 Gyr)
and young (2 Gyr) SSPs for changing metallicity, and for metal-rich ([Fe/H]=0) and metall-poor ([Fe/H]=-1.7) SSPs
for changing age, respectively.
Generally, the absolute differences between 8 and 4 Gyr old SSPs are larger than the differences between 8 and 12
Gyr old SSPs at fixed metallicity, as expected (Fig. \ref{abb.sensitivities}, lower panels). However, this effect
is much stronger at low than at high metallicity, which confirms what we get from the S parameter.
The absolute differences between models with different metallicity (top panels in Fig. \ref{abb.sensitivities})
are slighty larger for old than for than for young SSPs.
Interestingly, the plots show that indices known to be sensitive to age can also be highly variable for differing
metallicities; especially the broad Balmer indices H$\delta_A$ and $H\gamma_A$ change strongly with metallicity.
Most important, however, the plot shows that in practice, moderately metal-sensitive indices like Mg\textsl{b}
can be much more useful for metallicity determinations than indices like Fe5709 or Fe5782, though the latter are,
according to the S parameter, much more metal-sensitive.

In order to determine ages and metallicities of GCs, indices should be chosen not only according to known
sensitivities as given by S, but also according the achievable measurement accuracy and, if possible, according to
the expected age and metallicity range of the sources.

\section{The Lick Index Analysis Tool}
\label{Analysis}

Since in the original models (cf. Sect. \ref{synthesis}), the steps in metallicity are large, in a first step we
linearly interpolate in [Fe/H] between the 6 metallicities before we analyse any data with our new tool.
This is done in steps of [Fe/H] = 0.1 dex, so the final input grid for the analysis algorithm consists of sets of
all 25 Lick indices each for 28 metallicities ($-2.3 \le$ [Fe/H] $\le +0.4$) and 4000 ages (4 Myr $\le$ age $\le$
16 Gyr). Although this approach is only an approximation, the results shown in Sect. \ref{tests} prove it to be
sufficiently accurate.

\subsection{The $\chi^2$ - approach}
\label{chisquare}
The algorithm is based on the SED Analysis Tool presented by Anders \etal (2004); the reader is referred to this
paper for additional information about the algorithm, as well as for extensive tests using broad-band colors
instead of indices.\\

All observed cluster indices at once -- or an arbitrary subsample of them -- are compared with the models by
assigning a probability $p(n)$ to each model grid point (i.e., to each set of 25 indices defined by 1 age and 1
metallicity):
\begin{equation}
   \label{prob}
   p(n) \propto (-\chi^2) \ ,
\end{equation}
where
\begin{equation}
   \label{chi}
   \chi^2=\sum_{i=1}^{25}{\frac{(I_{\rm obs}-I_{\rm model})^2}{\sigma^2_{\rm obs}+\sigma^2_{\rm model}}}
\end{equation}
with $I_{\rm obs}$ and $I_{\rm model}$ being the observed and the model indices, respectively, and $\sigma_{\rm
obs}$ and $\sigma_{\rm model}$ being the respective uncertainties. Indices measured in magnitudes are transformed
into \AA ngstr\"om before calculation.
After normalization $(\sum{p(n)} = 1)$, the grid point with the highest probability is assumed to be the \emph{best
model}, i.e. it gives the ``best age'' and the ``best metallicity'' for the observed cluster.\\

The uncertainties of the best model in terms of $\pm 1 \sigma$ \emph{confidence intervals} are computed by
rearranging the model grid points by order of decreasing probabilities, and summing up their probabilities until
$\sum{p(n)} = 0.68$ is reached; the $1 \sigma$ uncertainties in age and metallicity are computed from the age
and metallicity differences, respectively, of the $n(p_{0.68})$- and the $n(p_{max})$-model. Note that the
determination does not take into account the possible existence of several solution ``islands'' for one cluster;
thus the confidence intervals are in fact upper limits.

\subsection{Examples and tests I: Galactic GCs}
\label{tests}
\begin{figure*}
   \begin{center}
   \includegraphics[width=0.34\linewidth,angle=270]{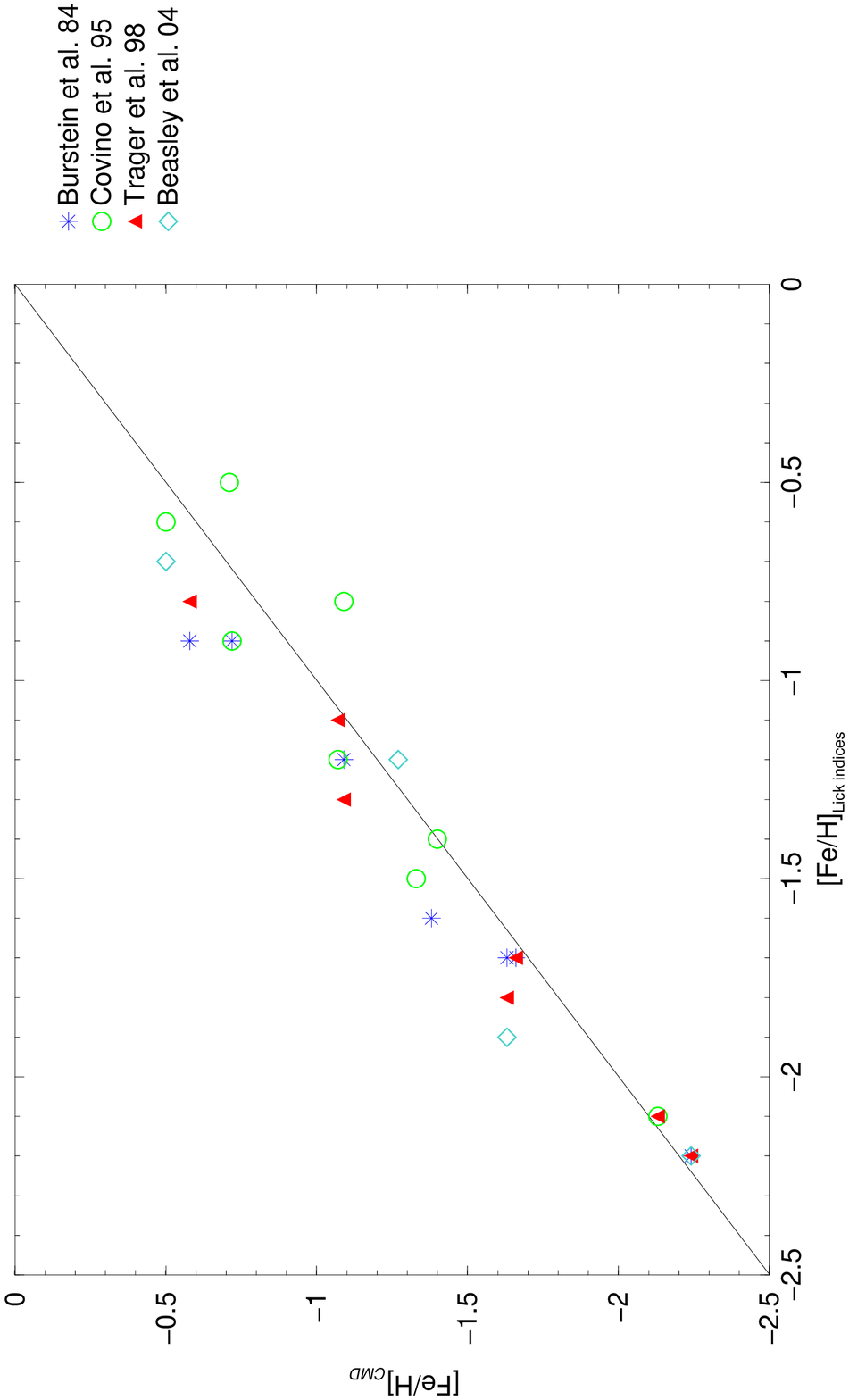}
   \includegraphics[width=0.34\linewidth,angle=270]{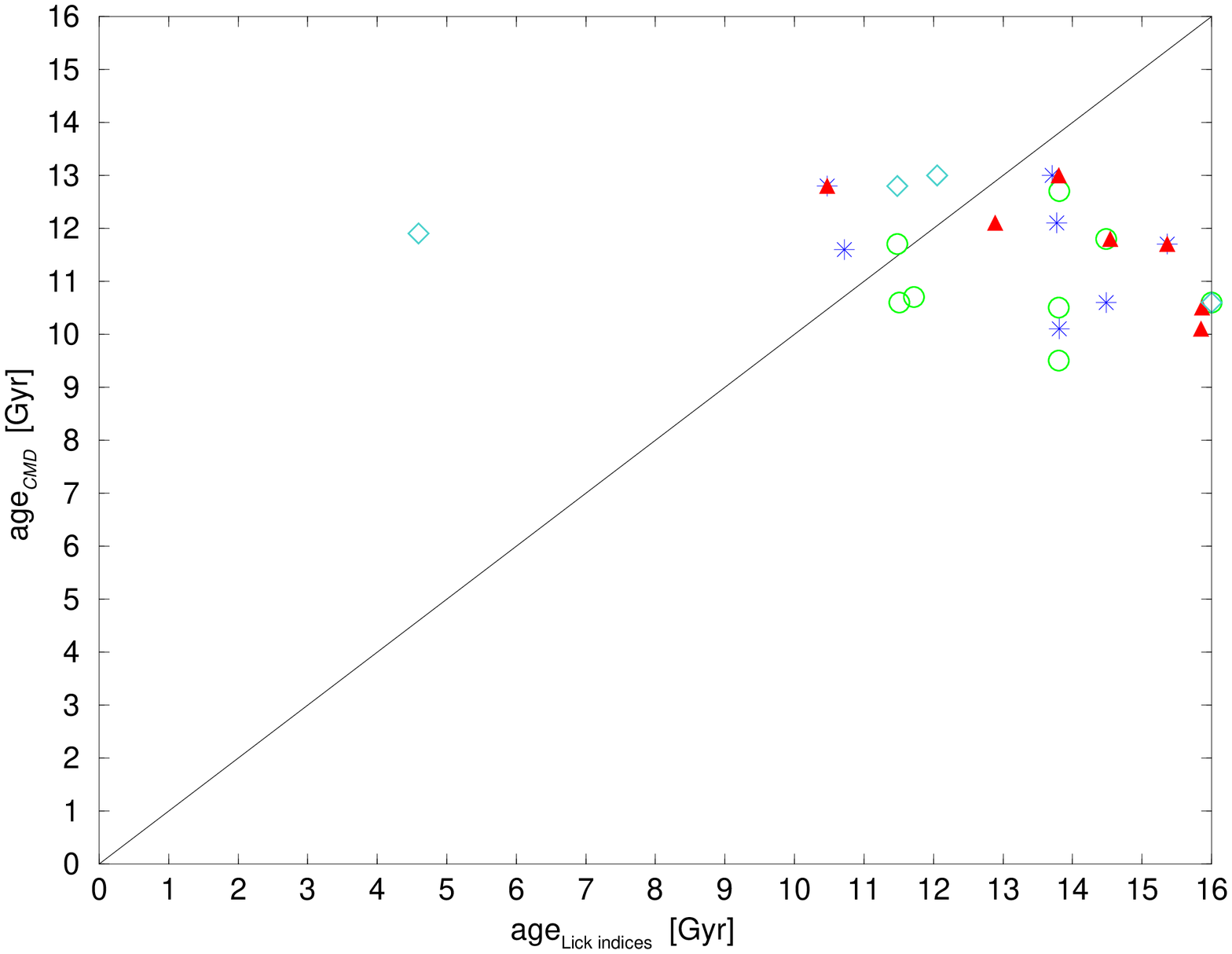}
   \caption{Galactic GC observations: Metallicities (\emph{left}) and ages (\emph{right}) determined using our Lick
   Index Analysis Tool (x-axis, using all measured indices available) vs. metallicities and ages determined by CMD
   analyses (y-axis, taken from Salaris \& Weiss 2002). Note that only results with \emph{model} uncertainties of
   $\sigma$(age) $\le$ 5 Gyr are plotted.}
   \label{abb.MWGC_allindices}
   \end{center}
\end{figure*}
\begin{figure*}
   \begin{center}
   \includegraphics[width=0.34\linewidth,angle=270]{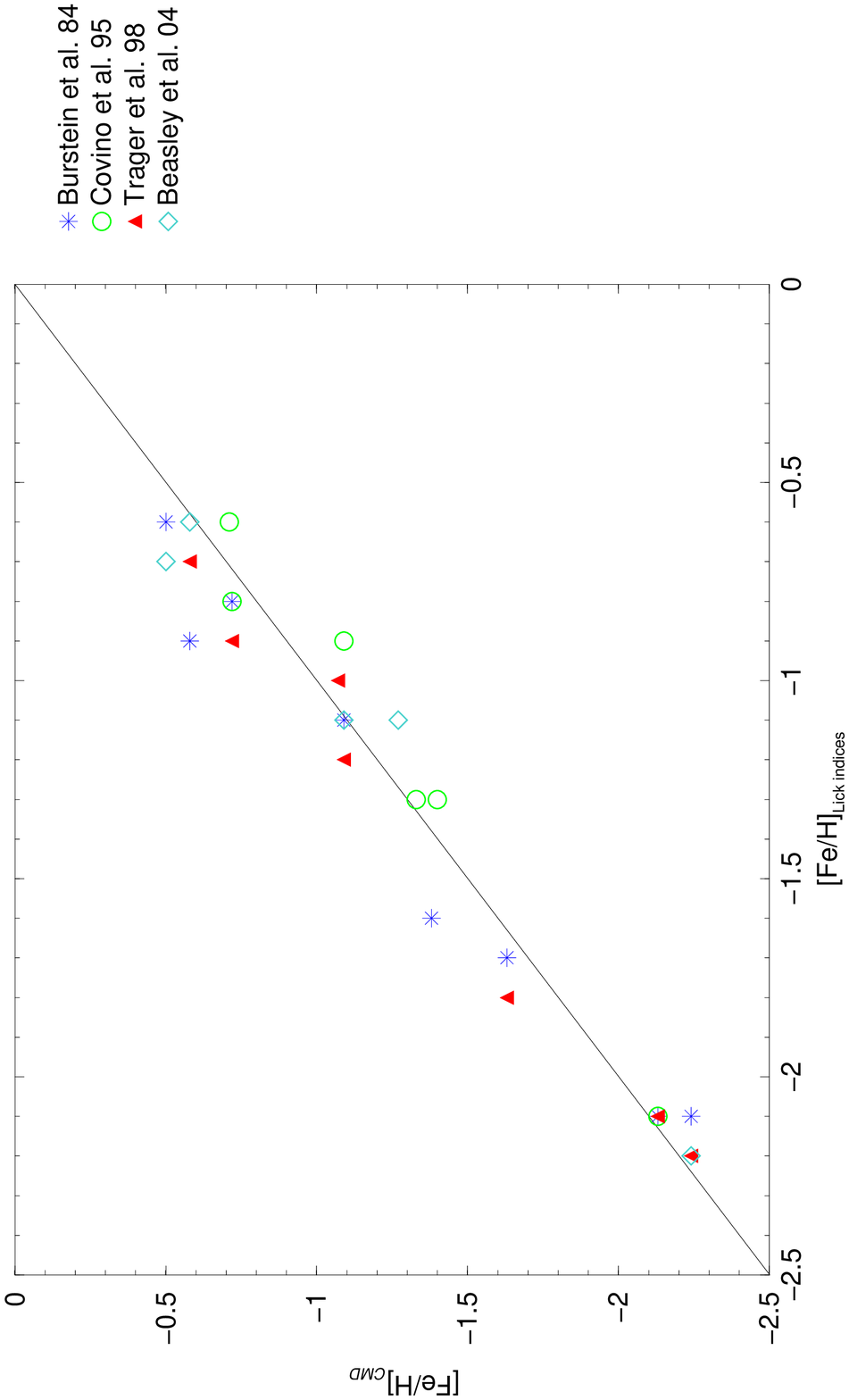}
   \includegraphics[width=0.34\linewidth,angle=270]{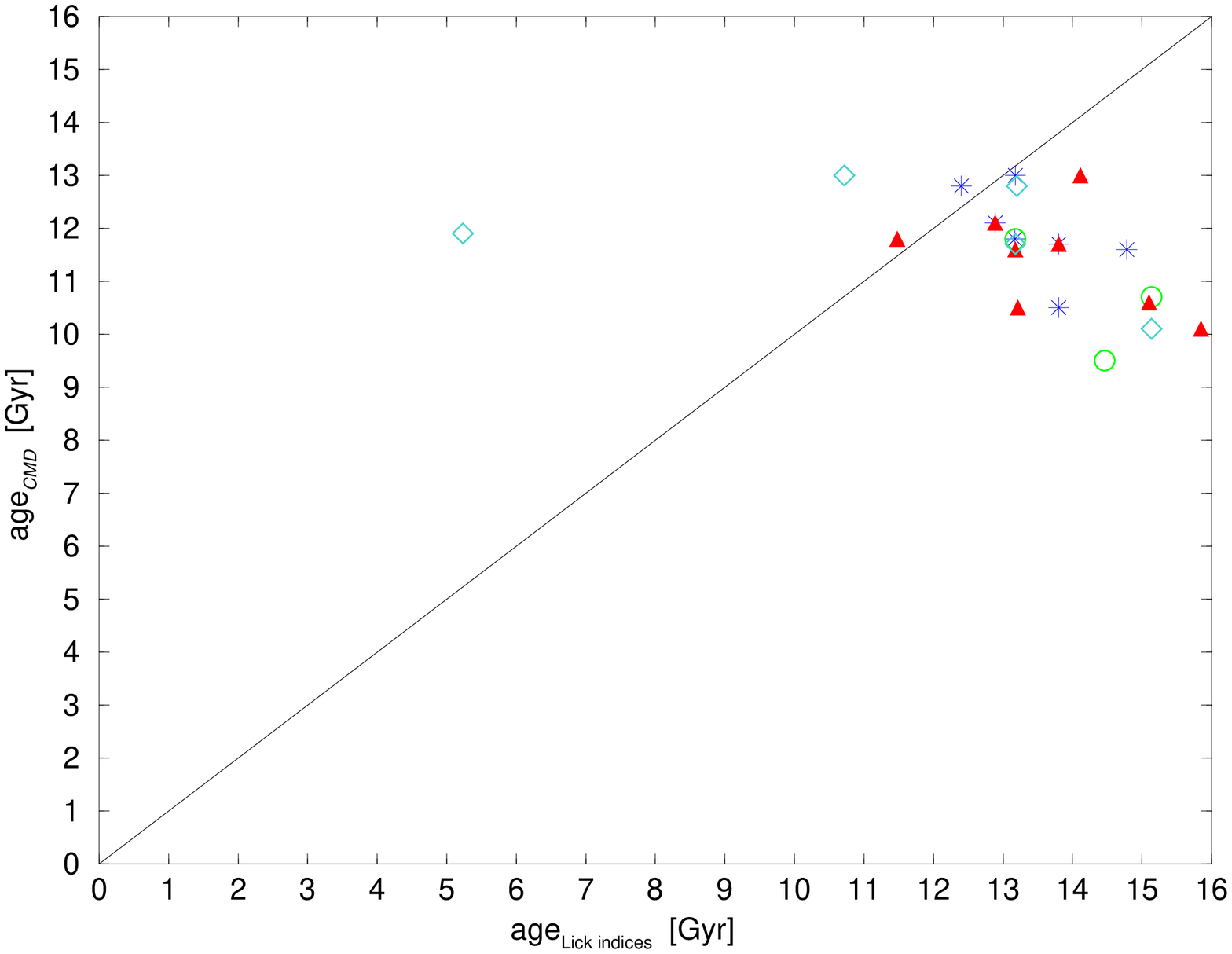}
   \caption{Same as Figure \ref{abb.MWGC_allindices}, but using metallicity-sensitive indices Mg$_1$, NaD, [MgFe]
   plus age-sensitive index H$\beta$ as input only (\emph{left}), and using age-sensitive indices Ca4227, G4300,
   H$\beta$, and TiO$_1$ as input only (\emph{right}).}
   \label{abb.MWGC_zandageindices}
   \end{center}
\end{figure*}

\begin{table}
   \caption{Observations by Burstein \etal (1984, B84), Covinio \etal (1995, C95), Trager \etal (1998,
   T98; H$\delta$, H$\gamma$ are taken from Kuntschner \etal 2002, see text),
   Beasley et al. (2004, B04) used to perform the tests of Section \ref{tests}.
   `*': index observed, `o': only a subsample of clusters is observed in this index.}
   \label{table.observations}
\centering
\begin{tabular}{c c c c c c}
        \hline\hline
                   & B84 & C95 & T98 & B04\\
        \hline
	CN$_1$		& * & o & * & o \\
	CN$_2$		&   &   & * & o \\
	Ca4227		&   &   & * & o \\
	G4300		& * & o & * & o \\
	Fe4383		&   &   & o & o \\
	Ca4455		&   &   & * & o \\
	Fe4531		&   &   & * & o \\
	Fe4668		&   &   & * & * \\
	H$\beta$	& * & * & * & * \\
	Fe5015		&   &   & * & * \\
	Mg$_1$		& * & * & * & * \\
	Mg$_2$		& * & * & * & * \\
	Mg\textsl{b}	& * & * & * & * \\
	Fe5270		& * & * & * & * \\
	Fe5335		& * & * & * & * \\
	Fe5406		&   &   & * & * \\
	Fe5709		&   &   & * & * \\
	Fe5782		&   &   & o & * \\
	Na D		& * & o & * & * \\
	TiO$_1$		& * &   & * & * \\
	TiO$_2$		& o &   & o &  \\
	H$\delta_A$	&   &   & * & o \\
	H$\gamma_A$	&   &   & o & o \\
	H$\delta_F$	&   &   & * & o \\
	H$\gamma_F$	&   &   & * & o \\
        \hline
\end{tabular}
\end{table}
We have tested our Lick Index Analysis Tool using a large set of Galactic GCs for which index measurements (taken
from Burstein \etal 1984, Covino \etal 1995, Trager \etal 1998\footnote{In this dataset, H$\delta_A$, H$\gamma_A$,
H$\delta_F$ and H$\gamma_F$ are taken from Kuntschner \etal 2002 who reanalysed Trager et al.'s spectra; in the
following, 'Trager \etal 1998' always is meant to include this additional data.}, and Beasley \etal 2004) as well
as age and metallicity determinations from CMD analyses (taken from Salaris \& Weiss 2002) are available.\\

Fig. \ref{abb.MWGC_allindices} compares ages and metallicities from both methods. Here, we use the complete set of
measured indices available (cf. Table \ref{table.observations}) as input for our Analysis Tool; for comparision,
Fig. \ref{abb.MWGC_zandageindices} shows our results using two subsets of indices: the age-sensitive indices
Ca4227, G4300, H$\beta$ and TiO$_1$ on the left panel, and metal-sensitive indices Mg$_1$, NaD, [MgFe] plus the
age-sensitive index H$\beta$ on the right panel\footnote{[MgFe] is a combination of metal-sensitive indices which
is known to be widely unaffected by non-solar abundance ratios (see, e.g., Thomas \etal 2003). It is defined as
[MgFe] := $\sqrt{\text{$<$Fe$>$} \times \text{Mg} b}$, with $\text{$<$Fe$>$} :=
\left(\text{Fe}5270+\text{Fe}5335\right) / 2$.}.
In all plots, only results with confidence intervals of $\sigma$(age) $\le$ 5 Gyr are plotted\footnote{In most
cases, very large 1$\sigma$ uncertainties are due to the presence of two ``solution islands'' (e.g., solution 1:
low or intermediate age, solution 2: high age) which are both within their 1$\sigma$ ranges. Since we do not want
to use any a priori information about the clusters, we cannot decide between the two solutions and therefore
rather omit them completely.}.

\begin{table}
   \caption{Mean ages and standard deviations of cluster ages determined using the Lick Index Analysis Tool and CMD
   analysis (Salaris \& Weiss 2002), respectively, as shown in Figs. \ref{abb.MWGC_allindices} and
   \ref{abb.MWGC_zandageindices}. Note that the values are computed without cluster NGC 6121.}
   \label{table.meanages}
\centering
\begin{tabular}{c c c c c}
        \hline\hline
                       & \multicolumn{2}{c}{all indices} & \multicolumn{2}{c}{age-sensitive indices}\\
                       & $<$age$>$ & $\pm$ & $<$age$>$ & $\pm$\\
        \hline
         Lick--Analysis & 13.49     & 1.80  & 13.58     & 1.22\\
         CMD--Analysis  & 11.54     & 1.08  & 11.57     & 1.04\\
        \hline
\end{tabular}

\end{table}

The agreement between [Fe/H] obtained from our Lick Index Analysis Tool and the corresponding values from CMD
analyses is very good, with $\Delta$[Fe/H] $\le$ 0.3 dex when using all available indices, and $\Delta$[Fe/H]
$\le$ 0.2 dex when using mainly metal-sensitive indices. With one exception, the age determinations are relatively
homogeneous, though the mean age obtained from index analyses is about 2 Gyrs too high compared to the results
from CMD analyses.
Table \ref{table.meanages} gives the mean ages and standard deviations of clusters determined
using the Lick Index Analysis Tool and from CMD analyses, respectively. It shows that, using all available
indices, not only the mean ages but also the age spreads are too high; most likely, this is due to varying
Horizontal Branch (HB) morphologies (see below). However, if only age-sensitive indices are used, the age spread is
of the same magnitude than that obtained by CMD analyses.

\begin{figure*}
   \begin{center}
   \includegraphics[width=0.4\linewidth,angle=270]{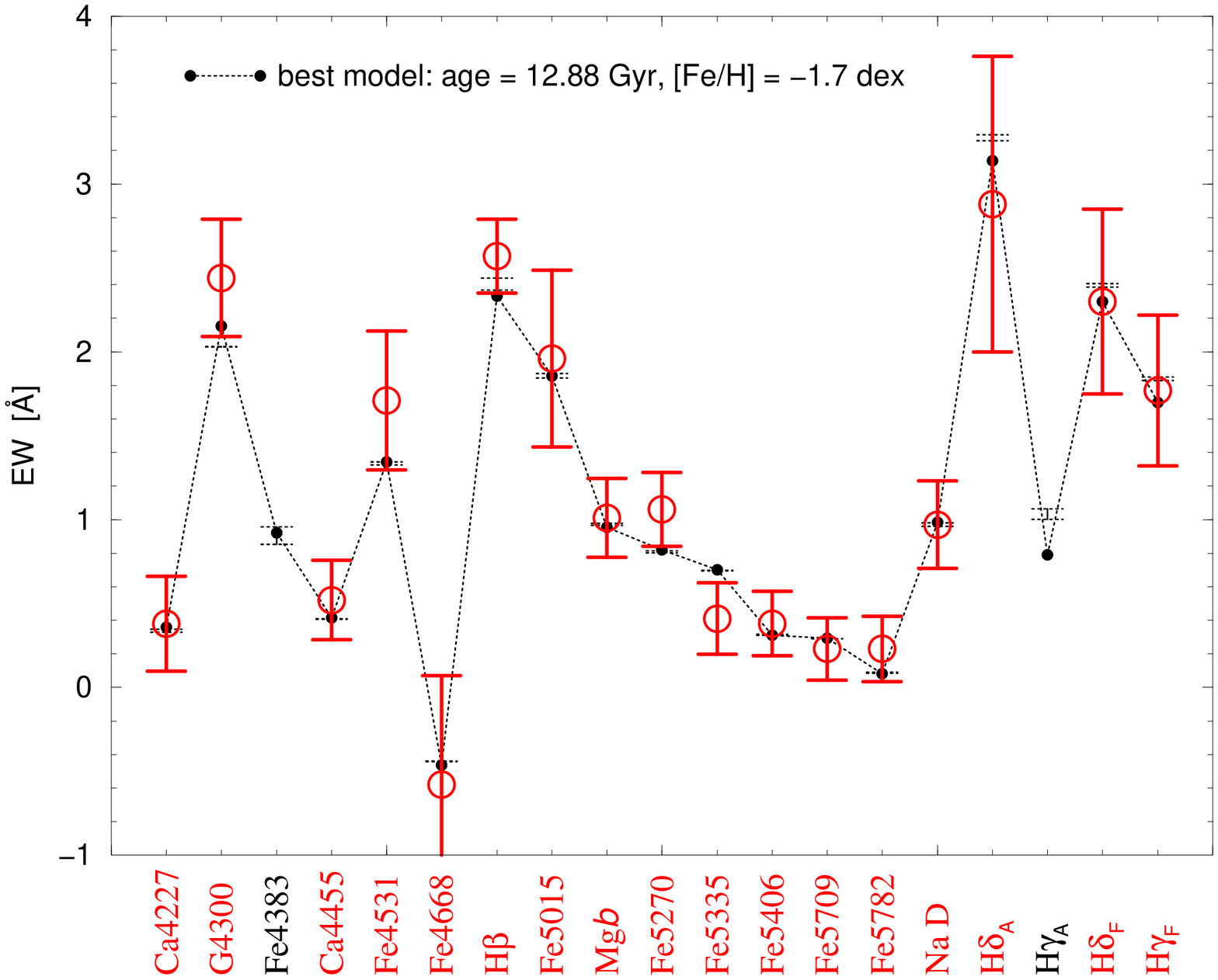}
   \includegraphics[width=0.4\linewidth,angle=270]{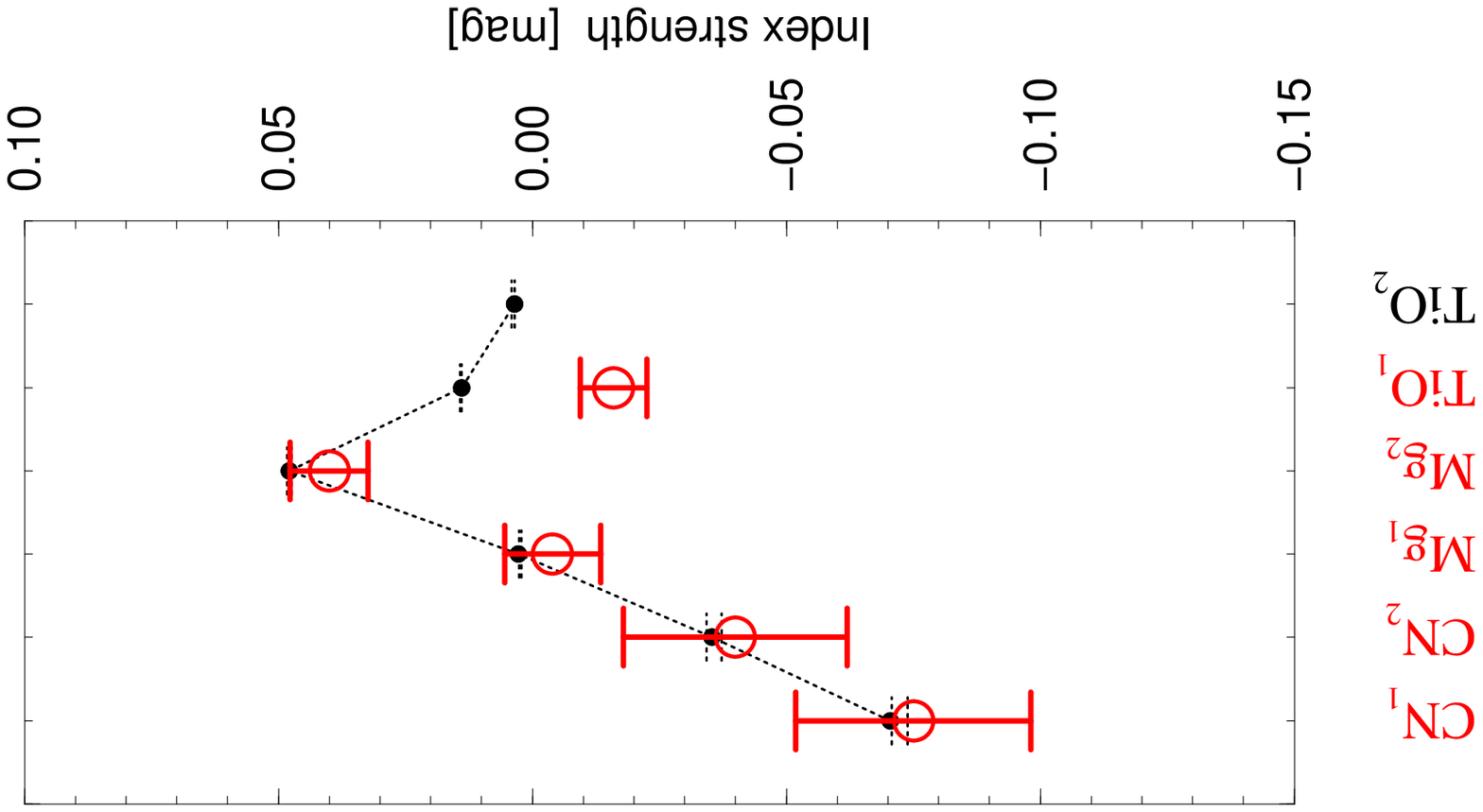}
   \caption{
   Lick index measurements of the Galactic GC M3 (NGC 5272) by Trager \etal (1998) with observational errors (open
   circles), and ``best model'' indices with the $\pm 1\sigma$ confidence intervals (black dots).
   The best model has an age of 12.88$(^{-1.99}_{+1.75})$ Gyr and [Fe/H] $=-1.7(\pm0)$ dex;
   Salaris \& Weiss (2002) give age = 12.1$(\pm0.7)$ Gyr and [Fe/H] $=-1.66$ dex.}
   \label{abb.n5272}
   \end{center}
\end{figure*}

\begin{figure*}
   \begin{center}
   \includegraphics[width=0.4\linewidth,angle=270]{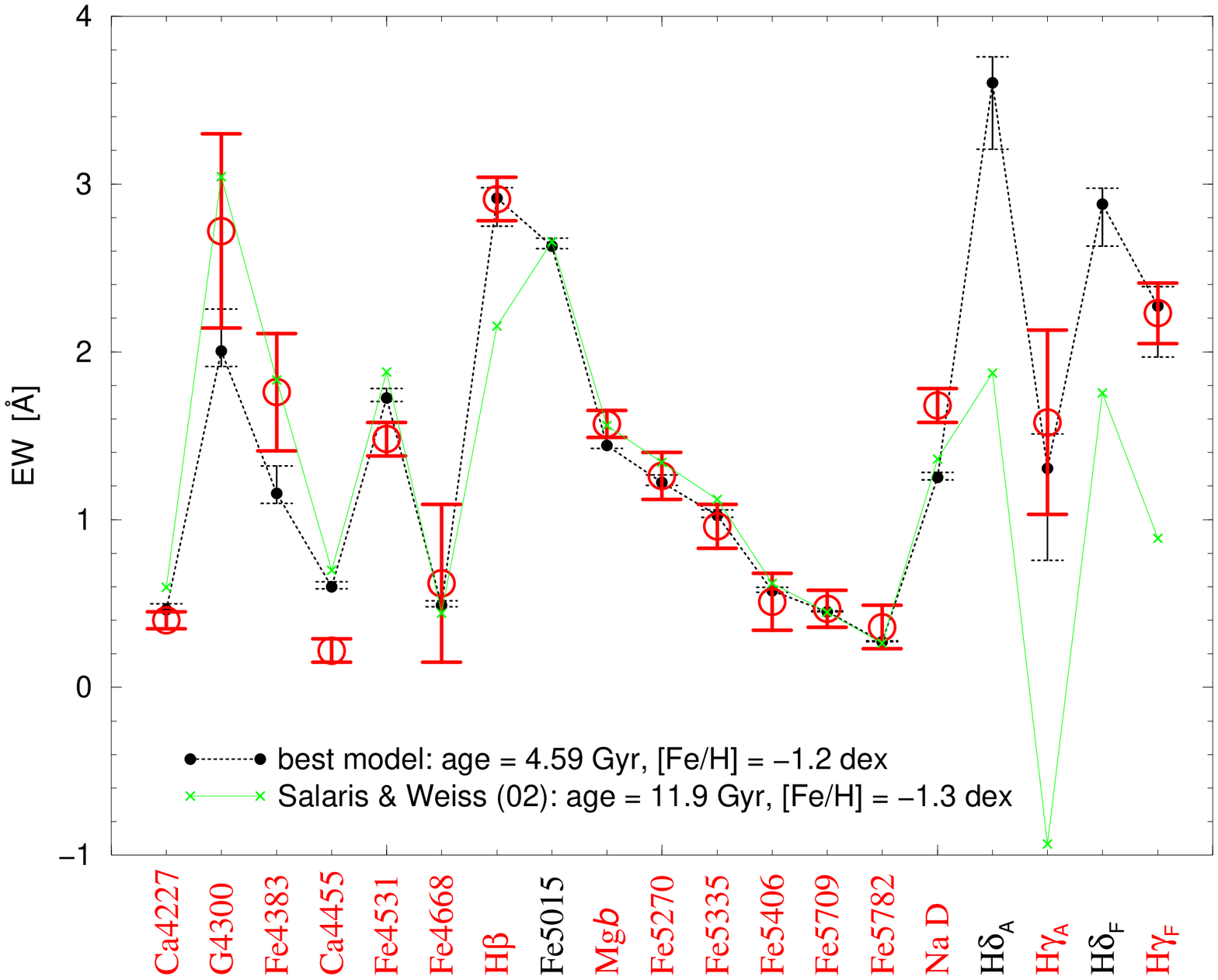}
   \includegraphics[width=0.4\linewidth,angle=270]{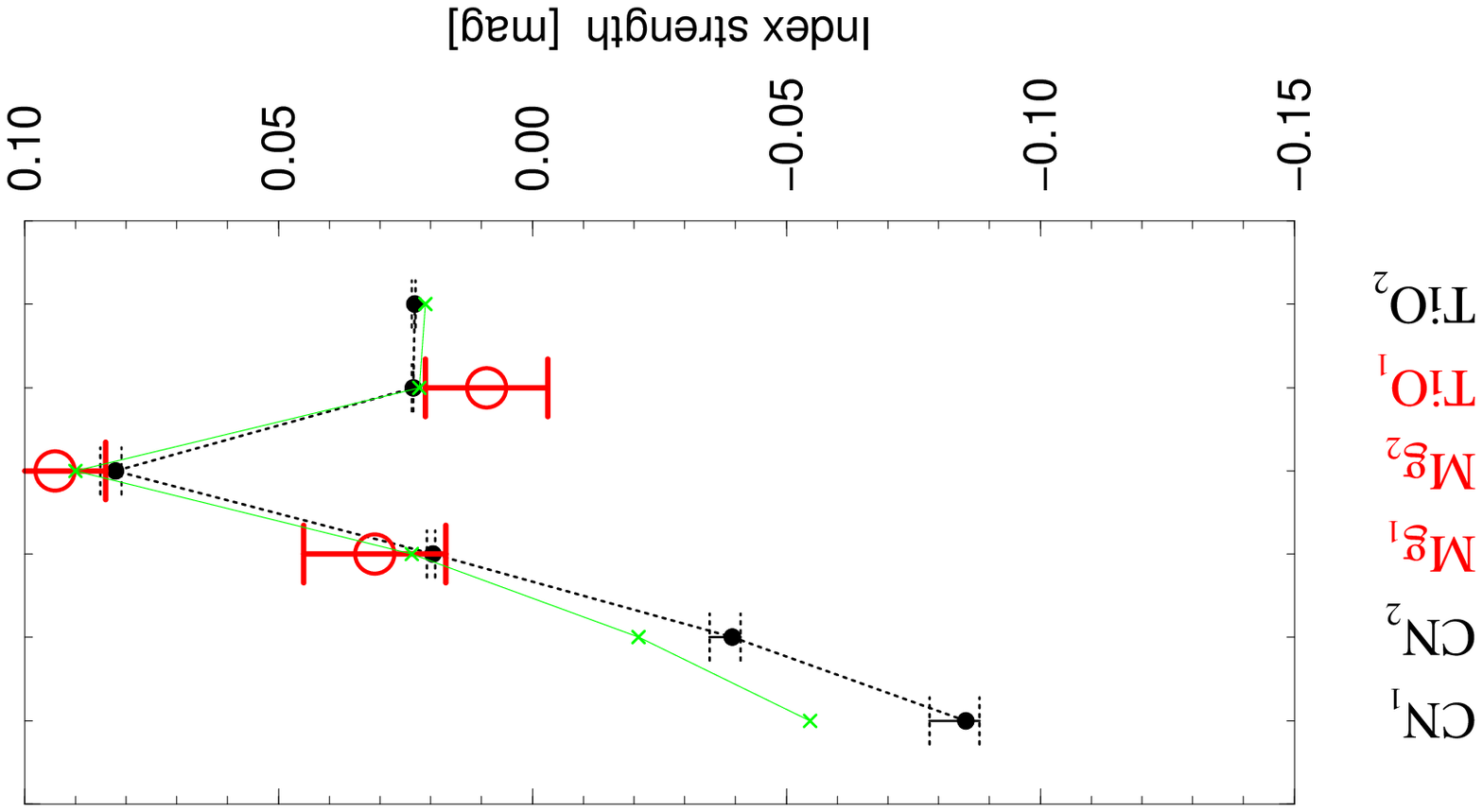}
   \caption{
   Lick index measurements of the Galactic GC M4 (NGC 6121) by Beasley \etal (2004) with observational errors (open
   circles), and ``best model'' indices with the $\pm 1\sigma$ confidence intervals (black dots).
   The best model has an age of only $4.59(^{-0.31}_{+0.80})$ Gyr and [Fe/H] $=-1.2(^{-0.1}_{+0.0})$ dex;
   Salaris \& Weiss (2002) give age = $11.9(\pm1.1)$ Gyr and [Fe/H] $=-1.27$ dex.
   Additionally, we plot model indices for the Salaris \& Weiss (2002) solution, i.e., a 11.9 Gyr / [Fe/H] $=-1.3$
   dex SSP model (small crosses).}
   \label{abb.n6121}
   \end{center}
\end{figure*}

As an example, Figure \ref{abb.n5272} shows the ``best-fitting'' model for the Galactic GC \object{M3}
(\object{NGC 5272}) together with the index measurements of Trager \etal (1998) used for the analysis. The best
model has an age of $12.88(^{-1.99}_{+1.75})$ Gyr and a metallicity of [Fe/H] $=-1.7(\pm0)$ dex; compared with an
age of 12.1$(\pm0.7)$ Gyr and [Fe/H] $=-1.66$ dex given by CMD analysis, this is a very good solution. Note that
the dotted line connecting the model indices is just for presentation.
We also give the $\pm 1\sigma$ confidence intervals of our best model in terms of index values for SSPs with age
$12.89-1.99=10.90$ Gyr and $12.89+1.74=14.63$ Gyr, respectively, and metallicity [Fe/H] $=-1.7$.

As seen in Figs. \ref{abb.MWGC_allindices} and \ref{abb.MWGC_zandageindices}, most of Galactic GCs are very well
recovered in their metallicities by our Lick Index Analysis Tool, in particular when the analysis is concentrated
on the metal-sensitive indices Mg$_1$, NaD, [MgFe] and the age-sensitive index H$\beta$. The origin of the $\sim$
2 Gyr systematic difference between index-determined and CMD-based ages, as well as of the wider age spread we
find is, most likely, due to the HB morphologies of the clusters:
The Padova isochrones we use for the analyses have very \emph{red} HBs over most of the parameter space (they have
blue HBs only for metallicities [Fe/H] $\le -1.7$ and ages higher than about 12 Gyr); therefore, the age of an
observed cluster with \emph{blue} HB can possibly be underestimated by several Gyrs.
Proctor \etal (2004), who use a similar technique than that applied here, also find ages too high
compared to values from CMD analyses; depending on the applied SSP models, they find mean ages of
13.1$(\pm2.3)$, 12.2$(\pm3.3)$ and 12.7$(\pm1.9)$ Gyr, respectively (cf. with Tab. \ref{table.meanages}).
We plan to analyse the influence of HB morphology on Lick index based age determinations in a separate paper.

Interestingly, and despite the fact that the Lick index measurements used here are of very different age and
quality, the results are of comparable quality for each data set.
E.g., the indices taken from Trager \etal (1998) are measured using the same original Lick-spectra than the
Burstein \etal (1984) data set; however, the spectra were recalibrated, and more indices were measured.
Nonetheless, the results from both data sets are comparable.\\

Though most results are acceptable, one cluster of our set is seriously misdetermined in terms of age: For the
Galactic GC \object{M4} (\object{NGC 6121}) the Lick Index Analysis Tool gives an age of only $\sim$ 5 Gyr (with a
1$\sigma$ uncertainty of less than 1 Gyr) both using all and only age-sensitive indices; CMD analysis gives more
than twice the age. Since the cluster does \emph{not} have a very blue HB (Harris 1996 gives a HB ratio of nearly
zero), we do not have a reasonable explanation for this.
However, anomalies have been found for this cluster, and some properties are still discussed in the literature
(see, e.g., Richer \etal 2004 and references therein).
Figure \ref{abb.n6121} shows models for this ``misdetermined'' cluster: Together with the index measurements
taken from Beasley et al. (2004), we show the index values for our best model (i.e., indices for a SSP with age =
$4.59(^{-0.31}_{+0.80})$ Gyr and [Fe/H] $=-1.2(^{-0.1}_{+0.0})$ dex) as well as for a model SSP using the Salaris
\& Weiss (2002) solution (age = $11.9(\pm1.1)$ Gyr and [Fe/H] $=-1.3$ dex).
The indices which differ most between the two models (and for which measurements are available) are G4300, Fe4383,
and the Balmer line indices H$\beta$ and H$\gamma$; remarkably, the Balmer lines seem to be completely responsible
for the misdetermination.

\subsection{Examples and tests II: M31 GCs and non-solar abundance ratios}
\label{alphatest}
\begin{figure*}
   \begin{center}
   \includegraphics[width=0.375\linewidth,angle=270]{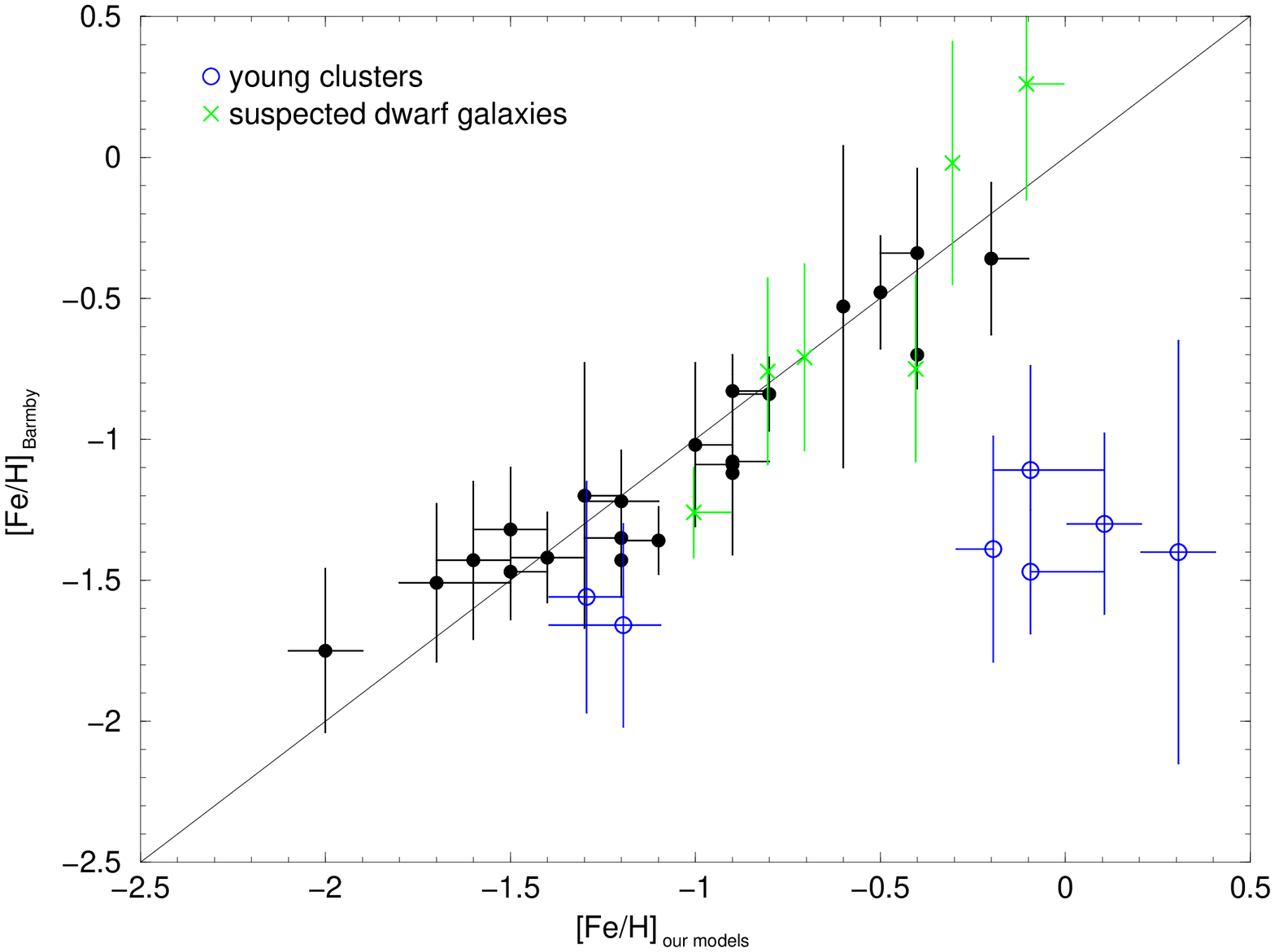}
   \includegraphics[width=0.375\linewidth,angle=270]{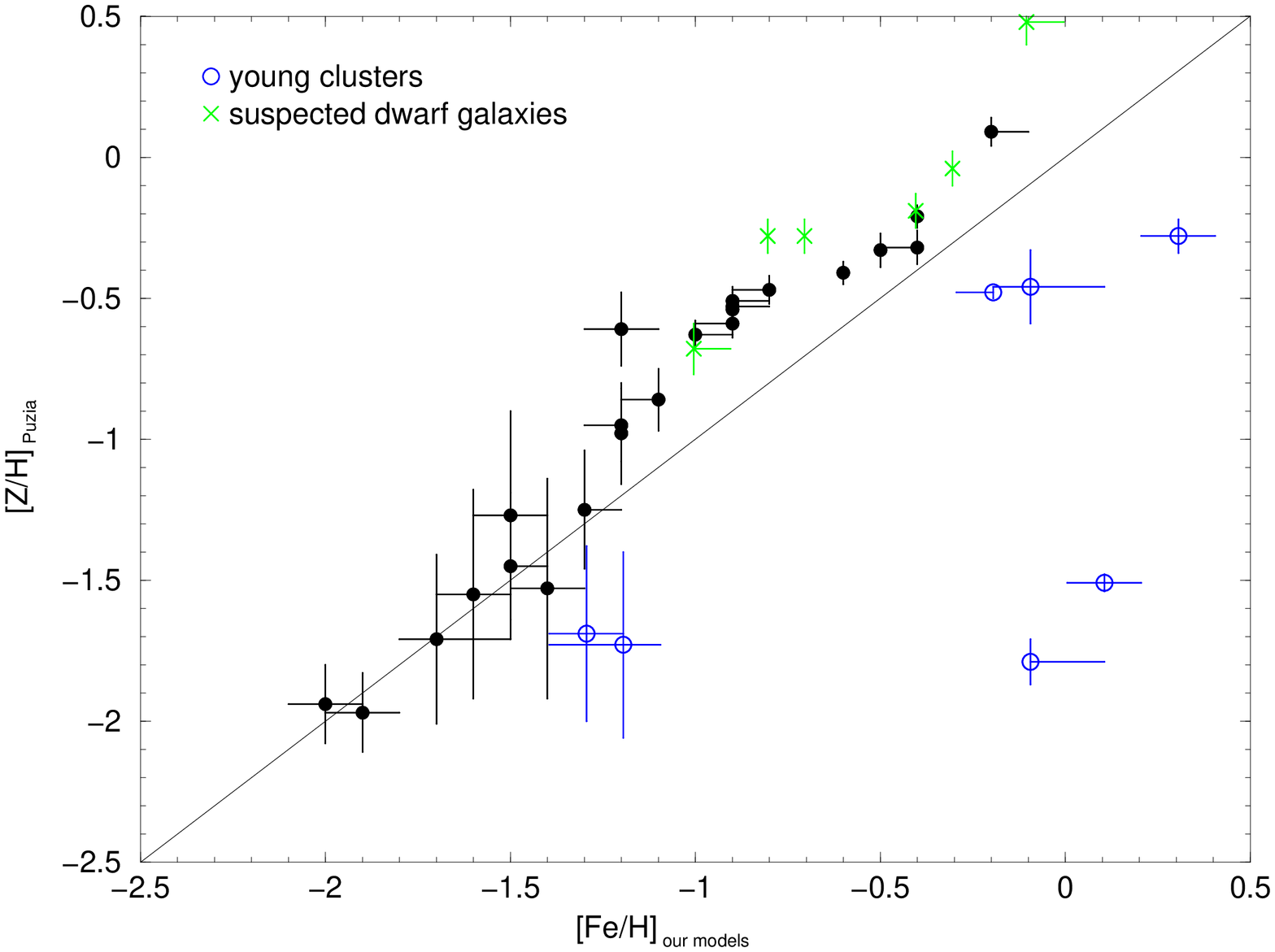}
   \includegraphics[width=0.375\linewidth,angle=270]{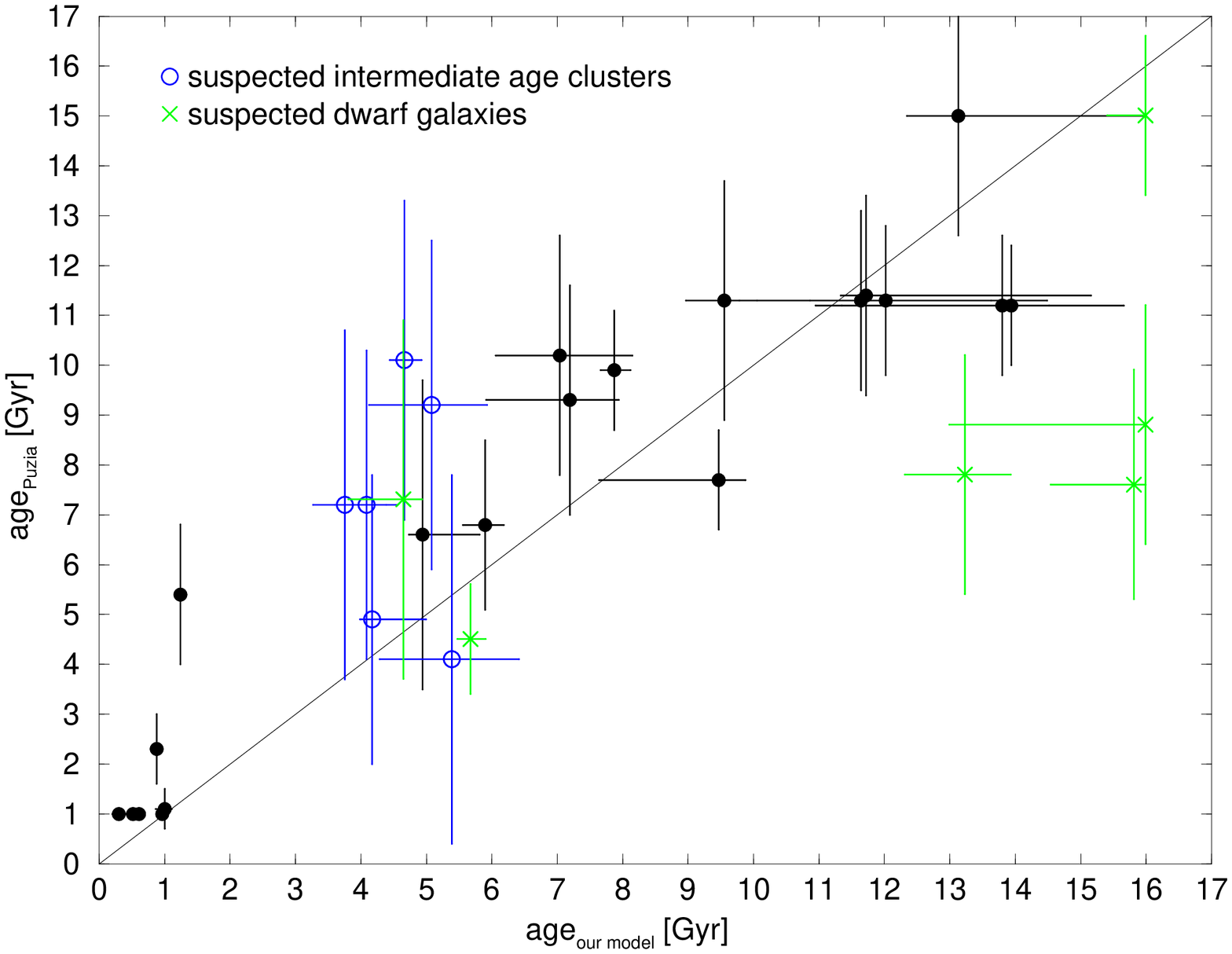}
   \caption{M31 GCs: Metallicities and ages for the Beasley \etal (2004) GC sample determined using our Lick
   Index Analysis Tool (x-axis, using all measured indices available) vs. metallicity determinations taken from
   Barmby \etal (2000) (\emph{top left}) and Puzia \etal (2005) (\emph{top right}), and vs. age determinations
   taken from Puzia \etal (2005) (\emph{bottom panel}). The classification as ``young cluster'' and ``suspected
   dwarf galaxy'' is taken from Beasley \etal (2004). See electronic edition for color versions of these plots.}
   \label{abb.M31_allindices}
   \end{center}
\end{figure*}
\begin{figure*}
   \begin{center}
   \includegraphics[width=0.333\linewidth,angle=270]{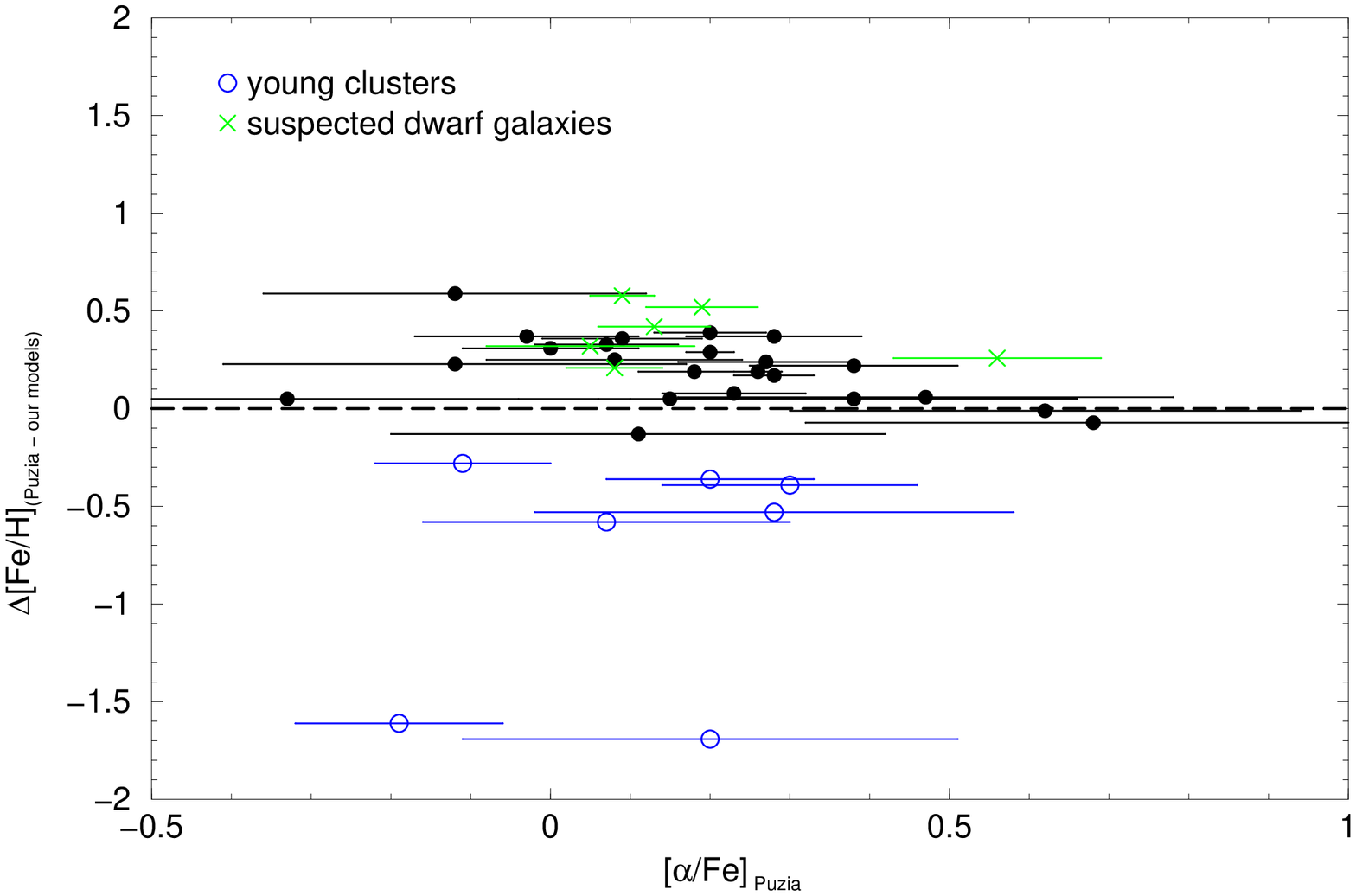}
   \includegraphics[width=0.333\linewidth,angle=270]{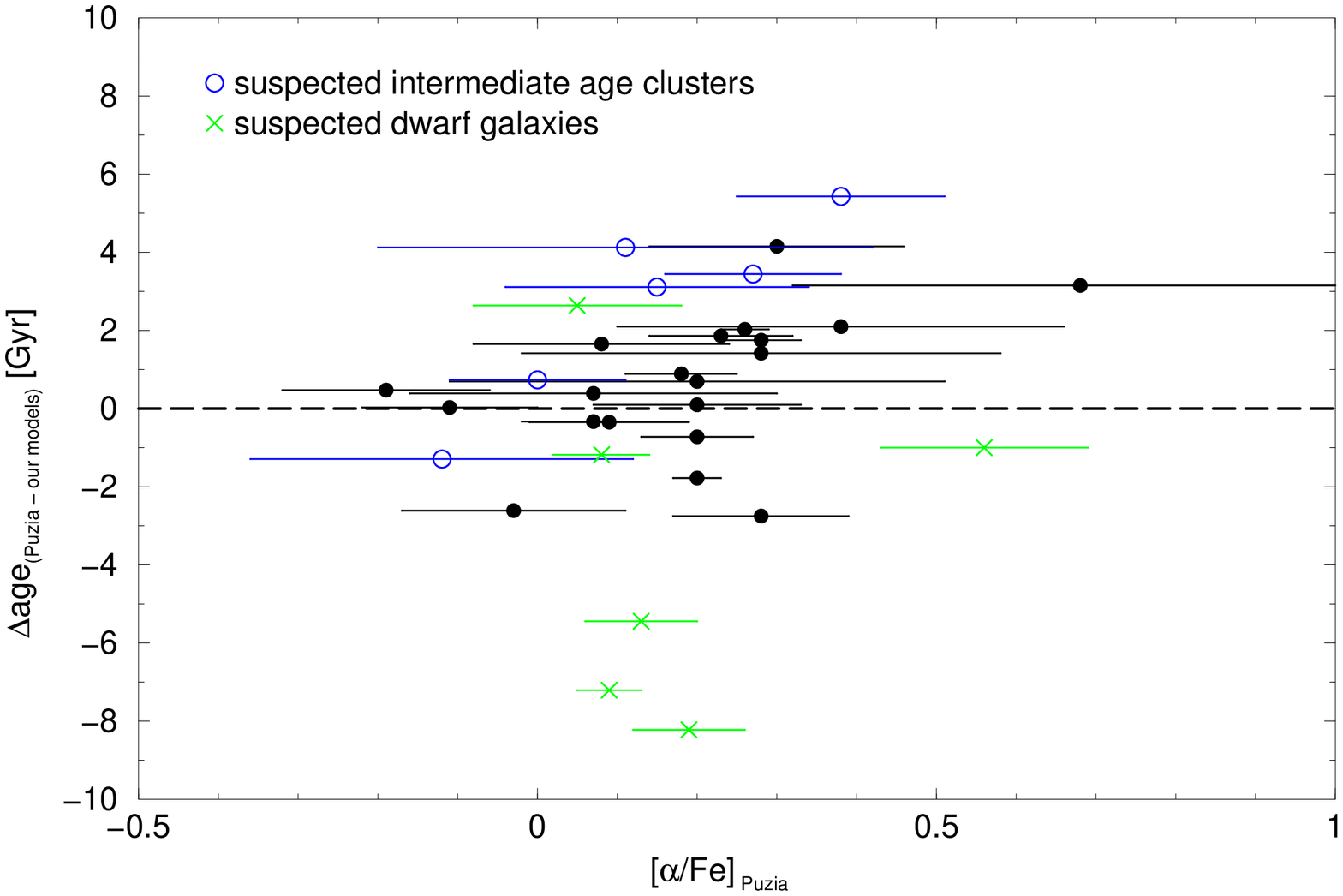}
   \caption{M31 GCs: Absolute differences between parameters derived using our Lick Index Analysis Tool and results
   from Puzia \etal (2005), against [$\alpha$/Fe] taken from Puzia \etal (2005).
   \emph{Left:} ([Z/H]$_{\text{Puzia}}$ $-$ [Fe/H]$_{\text{our models}}$);
   \emph{Right:} (age$_{\text{Puzia}}$ $-$ age$_{\text{our models}}$).
   The classification as ``intermediate-age cluster'' and ``suspected dwarf galaxy'' is taken from Beasley \etal
   (2004).}
   \label{abb.M31_delta_allindices}
   \end{center}
\end{figure*}

Unfortunately, for Andromeda galaxy (M31) GCs it is not possible to obtain high quality color magnitude diagrams;
reliable determinations of age and metallicity which could be used as ``default values'' for comparisions are not
available.
Therefore, for M31 GCs we can only compare our Lick index based determinations with results taken from the
literature which are based on spectral indices themselves.

For our analyses, we use the Lick index measurements of M31 GCs presented by Beasley \etal (2004); while not
presenting own age or metallicity determinations for individual clusters, they distinguish four classes for their
sample of cluster candidates: Young, intermediate age, and ``normal'' old GCs. Additionally, some sources are
suspected to be foreground galaxies. Beasley \etal have measured all available Lick indices with the exception of
TiO$_2$.\\

In Figure \ref{abb.M31_allindices}, we compare our metallicity determinations using the Lick Index Analysis Tool
with results presented by Barmby \etal (2000) (top left panel) and Puzia \etal (2005) (top right panel).
While Barmby \etal use calibrations given by Brodie \& Huchra (1990) for their spectroscopic metallicity
determinations, using own measurements of absorption line indices, Puzia \etal (2005) use a $\chi^2$ approach
using Lick index models from Thomas \etal (2003, 2004), which do account for non-solar abundance ratios. Puzia
\etal use the same database as we do (i.e., the Lick index measurements published by Beasley \etal 2004). They
give not [Fe/H] but total metallicities [Z/H]; however, according to Thomas \etal (2003), [Fe/H] in the ZW84
scale is in excellent agreement with [Z/H].
Hence, our results, given in [Fe/H], are perfectly comparable to Puzia et al.'s results, and well appropriate
to test for the influence of non-solar abundance ratios on our results.
For both the Barmby \etal (2000) and Puzia \etal (2005) metallicity determinations, we find good agreement with
our results. Only for clusters which are classified as young (i.e., age $\le$ 1-2 Gyr) we find relatively large
differences in [Fe/H]; however, this reflects our expectations, since the models are calibrated using
intermediate-age or old Galactic stars mainly.

In the bottom panel of Fig. \ref{abb.M31_allindices}, we confront our results with ages determined by Puzia \etal
(2005).
Again, the results are in surprisingly good agreement, if sources suspected to be foreground dwarf galaxies are
not considered. For the set of intermediate-age clusters identified by Beasley \etal (2004), our results
perfectly reflect this classification.

Compared with the classification of Beasley \etal (2004), the largest disagreements in both age and metallicity
occur for young clusters, and for suspected dwarf galaxies; it is no surprise that models computed to fit GCs are
not appropriate to be applied to galaxies (and, therefore, different methods lead to different results).\\

Since Puzia \etal (2005) also determine $\alpha$-enhancements for the GC sample, we can check for possible
systematic offsets of our determinations compared to theirs due to non-solar abundance ratios.

In Figure \ref{abb.M31_delta_allindices}, absolute differences between metallicities (left panel) and ages (right
panel) derived using our models and from Puzia \etal (2005) are plotted against [$\alpha$/Fe].
Relatively surprising, no general trend for the differences in both age and metallicity determinations with
$\alpha$-enhancement can be observed, if the large error bars of the [$\alpha$/Fe] determinations are taken into
account.
Hence, the slight offset between metallicities determined by Puzia \etal and by us (cf. Fig.
\ref{abb.M31_allindices}, top right panel) for [Fe/H] larger than $\sim -1.2$ dex seems \emph{not} to be due to
the use of solar-scaled against $\alpha$-enhanced models.

\section{Summary and outlook}
\label{Conclusions}
To cope with the observational progress that makes star cluster \& globular cluster spectra accessible in a large
variety of external galaxies,  we have computed a large grid of evolutionary synthesis models for simple stellar
populations, including 25 Lick/IDS indices using the empirical calibrations of Worthey \etal (1994) and Worthey \&
Ottaviani (1997). Comparison of the SSP models with Galactic GC observations shows good agreement between models
and data.

We find that the well-known and widely used age-sensitive indices H$\delta_A$ and H$\gamma_A$ also show a strong
metallicity dependence.
The ``metallicity sensitivity parameter'' S introduced by Worthey (1994) for old stellar populations with solar
metallicity is well reproduced by our models. Our models allow to extend this concept towards younger ages and
non-solar metallicities. We find the sensitivity of different indices with respect to age and metallicity to
depend itself on age and metallicity. E.g., all indices are generally more age sensitive at low than at high
metallicity. Another important issue is the \emph{absolute} difference in index strength for varying age or
metallicity: Due to the limited accuracy of any index measurement, these absolute differences in practice can be
of higher importance than the sensitivity given by S.

We present a new advanced tool for the interpretation of absorption line indices, the Lick Index Analysis Tool
\emph{LINO}. Following a $\chi^2$ - approach, this tool determines age and metallicity including their respective
$\pm 1 \sigma$ uncertainties using all, or any subset, of measured indices.
Testing our tool against index measurements from various authors for Galactic GCs, which have reliable age and
metallicity determinations from CMD analyses in the literature, shows very good agreement: Metallicities of GCs are
recovered to $\pm$ 0.2 dex using 6 appropriate indices only (Mg$_1$, Mg$b$, Fe5270, Fe5335, NaD, H$\beta$).
Age determinations from Lick indices consistently yield ages $\sim$ 2 Gyr higher than those obtained from CMDs.
The origin of this discrepancy is not yet understood.
Index measurements for M31 clusters are analysed and compared to results from the literature; a good agreement
between our results and age and metallicity determinations from the literature is found. We show that the
drawback of not having non-solar abundance ratio models do not seriously affect our results.

We will apply \emph{LINO} to the interpretation of intermediate-age and old GC populations in external galaxies,
complementing our SED Analysis Tool for the interpretation of broad-band spectral energy distributions.

All models are accessible from our website, \texttt{http://www.astro.physik.uni-goettingen.de/$\tilde{\
}$galev/}.

\acknowledgement
TL is partially supported by DFG grant Fr 916/11-1-2-3.\\


\end{document}